# Transport of water in a Transient, Impact-Generated Atmosphere on Mercury

J.K. Steckloff[1,2], D.B. Goldstein[1], P.L. Varghese[1], L.M. Trafton[3], P. Prem[4]

[1]Department of Aerospace Engineering & Engineering Mechanics, The University of Texas at Austin, 2617 Wichita Street, C0600, Austin, TX 78712-1221
[2]Planetary Science Institute, 1700 East Fort Lowell, Suite 106, Tucson, AZ 85719-2395
[3]Department of Astronomy, The University of Texas at Austin, 2515 Speedway, Austin, TX 78712-1205
[4]John Hopkins Applied Physics Laboratory, 11100 Johns Hopkins Road, Laurel, MD 20723

**Abstract**

Mercury's polar cold traps host water ice deposits that are likely populated with impact-delivered water via Mercury's exosphere. However, Mercury's near-sun location experiences an extremely high photodestruction rate that rapidly destroys water with a timescale of only ~3.5 hours. Here we use the PLANET DSMC code to investigate the fate of water from a single 1 km radius comet impact striking Mercury's North Pole (30 km/s at angle of 60°). We find that the evolving plume separates into four distinct phases: 1) an early plume phase in which ballistic escape and photodestruction reach their peaks, 2) a reentry phase in which water falling back toward the surface forms a self-shielded shock-topped atmosphere that migrates across the surface and ballistic loss ceases, 3) a quasi-steady phase in which a self-shielding dawn atmospheric enhancement (DAE) forms and drives, a tenuous migration of exospheric water to the cold traps with a longitudinal dependence, and finally 4) a late phase in which self-shielding ends and photodestruction dominates, effectively ending substantial water migration. In this work, we quantify the fates of the arriving water molecules, and describe some of the more important features of this highly unsteady, evolving three-dimensional atmosphere. We find that 23% of the initial water is photodestroyed, 65% of the water ballistically escapes the system (of which, 79% photodissociates prior to reaching the Hill radius), and 14% ends up in Mercury's cold traps, which is significantly more than the ~5% that migrates to the Moon's cold traps during an equivalent impact.

**Key Points**

- The evolving plume experiences four distinct phases: 1) an early phase, 2) a reentry phase, 3) a quasi-steady phase, and a 4) late phase.
- Of the initial water content of our comet, 23% is photodestroyed, 65% escapes ballistically, and 14% ends up in Mercury's cold traps.
- Self-shielding is essential for sufficient water to survive photodestruction to efficiently migrate to Mercury's cold traps.

**Plain Language Summary**

Mercury's poles have craters that contain large amounts of water ice. This water is thought to come from water-rich objects that strike Mercury's surface; vaporizing the water contained within, which can then migrate to these cold traps. However, some water molecules either escape the system, or are broken apart in Mercury's extreme near-sun environment. Here we use the PLANET DSMC code to compute the fates of the water contained within a single 1 km radius comet after it strikes Mercury's North pole at 30 km/s and an impact angle of 60°. We find that the evolving plume of water vapor can be divided into 4 distinct phases: an early plume phase in which the vapor plume expands, sweeping across the surface and expanding vertically, a reentry phase where the plume collapses back onto itself to produce a shock-confined atmosphere, a quasi-steady phase in which a self-shielding structure forms near the dawn terminator, and finally a late-phase where the atmosphere is extremely rarified. During these phases, we find that 23% of the initial water is broken apart by sunlight, 65% escapes back to space, and 14% ends up in Mercury's polar cold traps, where it is indefinitely stable against further thermal evolution.

## I. Introduction

Cold traps are locations on a planet's surface cold enough to accumulate/store water ice over geologic time. Typically, cold traps occupy the permanently shadowed regions (PSRs) on the floor of high latitude craters (*Watson et al. 1961; Arnold, 1979*) such as those at Mercury's poles, which contain numerous PSRs where water ice could stably collect (*Paige et al. 1992, 2013*). Water ice was first detected within Mercury's polar cold traps with Earth-based planetary radar (*Slade et al. 1992; Harmon & Slade 1992; Butler et al. 1993; Harmon et al. 1994, 2001, 2011*). The MErcury Surface, Space ENvironment, GEochemistry, and Ranging (MESSENGER) spacecraft later confirmed these detections with neutron spectrometry (*Lawrence et al. 2012*) and visible reflectance (*Neumann et al. 2013; Chabot et al. 2014, 2016*). Both Earth-based radar and *in situ* MESSENGER observations suggest that this ice was recently emplaced (*Butler et al. 1993; Ernst et al. 2018*) within the last ~150 Myr (*Deutsch et al. 2019*), consistent with this ice being delivered predominantly by a single large impact with a speed up to 30 km/s (*Ernst et al. 2018; Deutsch et al. 2019*), such as the large Hokusai impact event (*Ernst et al. 2018*). This is consistent with work that found that larger impacts are better able to populate Mercury's cold traps due to the large resulting water vapor plume, which shields itself from solar radiation flux that would otherwise rapidly photodissociate the $H_2O$ due to Mercury's proximity to the Sun (*Steckloff et al. 2022*).

Exospheric dynamics have been suspected to facilitate the migration of impact-delivered water to Mercury's cold traps (*Harmon & Slade 1992; Butler 1997; Ernst et al. 2018*), and numerous studies support this. Modeling based initially on theory and later MESSENGER observations reveal that water is stable within Mercury's PSRs over geologic timescales (*Paige* *et al 1992; Ingersoll et al. 1992; Vasavada et al. 1999; Paige et al. 2013*), and that impacts can deliver sufficiently large quantities of water to Mercury (*Moses et al. 1999*). Furthermore, Monte Carlo studies showed that simulated molecules scattered across Mercury's surface can indeed migrate across the surface (*Wurz and Lammer 2003*) to polar cold traps via free molecular flow (*Schorghofer 2015; Feoktistova et al. 2021*). Further studies explored the burial process of volatiles in these cold traps via space weathering processes (*Crider and Killen 2005*).

However, studies to date have yet to explore the full impact delivery process. This is partly of necessity, as it requires the ability to simulate the impact process, the resulting highly collisional impact-induced vapor plume, and the highly unsteady, fully 3D nature of the evolving plume including the coupling of the radiation transport and gas dynamics, which requires a code that can handle all these different processes. In this work, we make a first attempt at understanding how the vapor plume resulting from the impact of a 1 km radius comet nucleus into Mercury's North Pole evolves and migrates, and quantify the ultimate fate of these water molecules, and then compare with previous results for identical impacts into the Moon.

## II. Methods

To understand how volatile-rich impactors produce vapor plumes that form transient atmospheres and populate the cold traps of Mercury, we combine impact hydrodynamic simulations with exospheric evolution codes that use the direct simulation Monte Carlo (DSMC) method (*Bird 1994)*. We combine these simulations with an up-to-date map of Mercury's cold traps, to understand the cold trapping process. For this study, we used the SOVA (SOlid, Vapor, Air) hydrocode (*Shuvalov, 1999, Ivanov and Artemieva, 2002*) hydrodynamic simulations of a canonical 1 km radius comet striking the north pole of a planetary body at 30 km/s at a 60° impact angle, and then feed these results into the PLANET DSMC code to study the evolution of the resulting water vapor plume, a technique previously developed and used by *Stewart et al. (2011)* and *Prem et al. (2015)*.

We have already completed SOVA simulations of a 60°, 30 km/s comet impacting the Moon, and have already computed the first 30 s of the resulting vapor plume's evolution using PLANET DSMC. This is close to the average comet impact speed at Mercury (37 km/s; *Moses et al. 1999*), but is not the most probable impact angle (which is 45 degrees); regardless, this allows us to reuse these existing simulations, albeit with careful consideration of how the properties of Mercury would have affected this evolving plume. Such impacts into a large body such as Mercury will result in a global fallback of a portion of the ejected material (see section 3.2); thus the effects of using different impact speeds can be computed using ejecta scaling laws for impactor material (e.g., *Melosh 1989; Housen and Holsapple 2012; Hyodo and Genda 2020*), which we discuss later (see section 4.1).

### *2.1 Impact Modeling*

We reuse the impact simulations from *Prem et al. (2015)* who used the SOVA hydrodynamic shock physics code (*Shuvalov 1999)* to simulate the full three-dimensional impact and immediate aftermath of a 1 km radius comet striking at the north pole. Hydrodynamic shock physics codes, or "hydrocodes" use an equation of state to compute the continuum behavior of a material under situations of extreme pressure and temperature, conditions under which even rock behaves as a fluid for the processes under consideration; this makes hydrocodes ideal for describing the behavior of impactor and target materials throughout the impact process. As described in *Prem et al. (2015)*, the SOVA simulation modeled a comet as a 1 km radius sphere of water ice striking a dunite target surface at 30 km/s at a 60° angle (measured from the horizontal). For further details on these SOVA simulations, we refer the reader to *Stewart et al. (2011)* and *Prem et al. (2015)*.

Such hypervelocity impacts are sufficiently energetic to completely vaporize all water ice present in the simulated comet. Initially, the water vapor near the point of impact is sufficiently dense to be accurately described by SOVA's continuum solver. However, the vapor plume rapidly rarefies as it expands, limiting the accuracy of the SOVA computations for this vapor plume to the region within 20 km of the impact (*Prem et al. 2015*). At this point, the SOVA outputs are handed off to the PLANET DSMC code, which can compute the behavior of the rapidly rarefying plume.

### *2.2 PLANET DSMC*

As the expanding plume rarefies, continuum fluid mechanics analyses using the Navier-Stokes equations begin to break down and different aspects of radiative transfer not treated by SOVA start to become important; these conditions require the use of the Boltzmann equation and radiative transfer models to accurately describe the behavior of the vapor plume. Unfortunately, relevant analytic solutions to the Boltzmann equation do not exist; the direct simulation Monte Carlo (DSMC) technique avoids this complication by modeling the dynamical and collisional behavior of a representative sample of molecules probabilistically, from which the gas flow field is determined statistically (*Bird, 1994)*. The ability of DSMC techniques to simulate gas flows in moderately collisional atmospheres and exospheres has led to the adoption of the technique in planetary science.

To generate the representative sample of lofted water molecules, *Prem et al. (2015)* followed the method of *Stewart et al. (2011)*, which defines a thin hemispherical interface zone 20 km in radius and centered on the impact site. DSMC particles are generated by sampling from the material in the SOVA simulations as it crosses into this interface zone, in which both the SOVA and the DSMC grids overlap. Our DSMC simulations only consider water vapor; if a parcel of material is rock from the target (rather than the impactor), it is ignored and no DSMC particles are generated. As the impact process mixes together water from the comet impactor and rock from target material, some parcels of material are a mixture of both rock and water vapor; such mixtures are also ignored at the SOVA-DSMC interface zone, and no mixed particles are generated (*Stewart et al. 2011; Prem et al. 2015)*; previous studies of hybrid SOVA-DSMC simulations found that such mixed materials contain, at most, 3% of the total water vapor in the simulations (*Stewart et al. 2011*); this is acceptable for our study.

Descriptions of the capabilities, structure, and design of the PLANET DSMC code used for this work can be found in *Stewart et al. (2011), Prem et al. (2015),* and *Steckloff et al. (2022)*. Briefly, our PLANET DSMC code is a fully 3D DSMC solver, using $O(10^8)$ particles and $O(10^6)$ cells in the computational mesh of these simulations. PLANET uses a Variable Hard Sphere model (*Bird 1994*) for computing probability of collision between molecules. PLANET DSMC is optimized to run on large parallelized supercomputing clusters, typically using ~188 CPUs for our simulations, and are spread across multiple compute nodes. Due to resource and access constraints, simulations run for a maximum of 48 hours (wall clock time, not CPU hours), after which the code needs to record the state of the simulation, and be restarted. The amount of time that can be simulated in these 48 hours varies depending on the complexity, state, and computational requirements of the evolving simulated system. For our purposes, the simulation tends to evolve over time from a more complex, unsteady state to a simpler quasi-steady state; this tends to require fewer computational resources per unit of simulated time, allowing the amount of flow evolution simulated in 48 hours to grow steadily as the simulation progresses.

When a molecule strikes the surface, it is assumed to adsorb/condense onto the surface (*Langmuir 1916*), and we use the surface temperature-dependent residence time model of *Frenkel (1924)* to determine if the molecule desorbs/sublimes during a given timestep (*Stewart et al. 2011; Prem et al. 2015*). By default, PLANET determines the surface temperature of an airless body as a geometric function of the distance of the surface from the subsolar point:

$$T(\theta, \phi) = \begin{cases} T_{min} + (T_{peak} - T_{min})\left[\cos\theta \cos(\phi - \phi_0(t))\right]^{1/4} & \text{if } \left|\phi - \phi_0\right| \leq 90^\circ \\ T_{min} & \text{if } \left|\phi - \phi_0\right| > 90^\circ \end{cases} \quad (1)$$

where $T_{peak}$ and $T_{min}$ are the peak (i.e., subsolar) and minimum surface temperatures respectively, $\theta$ is latitude, and $\phi_0$ is the subsolar longitude. Non-zero values of $T_{min}$ cause deviations in temperature from an idealized surface in local radiative equilibrium; *Steckloff et al. (2022)* nevertheless found that this thermal model provides sufficiently good agreement for our purposes. For Mercury, $T_{min}$ = 100 K and $T_{max}$ = 700 K (*Vasavada et al. 1999*). Of note, PLANET DSMC assumes that the planet (here Mercury) has zero pole obliquity (a.k.a., axial tilt), which is close to Mercury's observed 2 °degree pole obliquity (*Margot et al. 2012*). We note that this surface temperature model assumes an optically thin frost layer.

PLANET DSMC uses values from *Huebner et al. (1992)* and *Huebner et al.* (*2015*) to compute rates of photodestruction at 1 AU of molecules exposed to sunlight (*Stewart et al. 2011; Prem et al. 2015*); these values are scaled up according to the inverse square law, to account for the increased solar intensity at Mercury's heliocentric distance. Because sunward molecules can absorb the UV photons responsible for photodestruction (and therefore shield the molecules behind them from photodestruction), PLANET DSMC includes a unique solver that computes column densities experienced by solar photons throughout the computational mesh, and attenuates the intensity of solar radiation and resulting photodestruction to account for this atmospheric self-shielding effect (*Prem et al. 2015*). Molecules are allowed to spontaneously lose energy by radiating from rotational and vibrational states (*Prem et al. 2015*). PLANET DSMC also includes a sophisticated treatment of radiative heat transport within the vapor plume. PLANET allows for photon packets emitted from the surface (i.e., Stefan-Boltzmann emission) and from photons emitted from the gas molecules themselves to propagate through the computational mesh in a Monte Carlo radiative transfer model that uses the gas density in each grid cell to compute how much energy from these packets of photons is deposited by gas absorption in each cell (*Prem et al. 2019a*); this radiative transfer model assumes that relative speeds between the emitting and absorbing molecules are negligible compared to the speed of light (negligible red/blue shift).

In our simulations we initially compute these radiation and thermal fields every 100 timesteps (initial timestep is 1s), or at the beginning of every simulation restart (whichever is shorter). The initial restarts occur after every 30-60 s of simulated time, before steadily growing longer than the 100 s radiation-transport compute cadence. This allows us to achieve reasonable time resolution when the plume is evolving most rapidly. After 200 hours, the plume has long since settled into a quasi-steady state, allowing us to compute the radiation and thermal field much less frequently (so long as the computation interval is small compared to the time needed for Mercury's surface to rotate through the arclength of a single grid cell). After 500 hours, we increase the timestep to 10 s, thus recomputing the radiation and thermal grid values every 1000 s.

This work assumes that the simulated body has a circular orbit and that the apparent motion of the Sun across the sky is uniform. Because PLANET DSMC models a day on a body by treating the body as fixed and modeling the Sun going around the object, a single specified rotation period is used to both model the motion of the Sun across the sky and to compute the non-inertial forces (e.g., Coriolis forces) that result from this non-inertial reference frame. This approach generally works well for all airless solar system bodies to which PLANET has modeled in the past: the Moon (*Stewart et al. 2011; Prem et al. 2015; 2019a*), Io (*Walker et al. 2010;2012; McDoniel et al. 2017; Ackley et al. 2021; Hoey et al. 2021*), Ceres (*Steckloff et al. 2022*), Enceladus, Pluto-Charon (*Hoey et al. 2017*). However, Mercury is unique among the planets in significantly deviating from a circular orbit (orbital eccentricity of 0.21) and having a spin-orbit coupling (3:2 spin-orbit resonance). As a result, the Sun tracks across the skies of Mercury at different rates, leading to two "hot poles" and two "cold poles". Nevertheless, we neglect these complications, as we are interested exclusively in how a generic impact-induced vapor plume on Mercury would evolve due to Mercury's slow rotation rate, relatively high gravity for an airless body, and proximity to the Sun. We therefore model Mercury as having a circular orbit with a radius of 0.387 AU and a uniform rotation period (sunrise-to-sunrise) of 88 days. This is longer than Mercury's sidereal day length (rotation period relative to the fixed stars) of 58.6 days, but is half the length of Mercury's synodic day length (sunrise-to-sunrise) of 176 days. The effect of this approximation is a slight error in the already-small non-inertial forces, and with material being fed into Mercury's resulting dawn atmospheric enhancement (DAE) at a rate twice as fast (see section 3.3.1). Nevertheless, these errors represent minor perturbations on the general trends that are the focus of this work, and are therefore acceptable. In separate work, we have since implemented the ability to model Mercury's orbit more accurately (*Prem et al. 2025*).

*2.3 Reusing Lunar SOVA-DSMC Simulations*

Once SOVA simulation results are fed into PLANET DSMC, the resulting vapor plume is hot and dense, requiring high spatial and temporal resolution in the DSMC simulations to keep the cell size comparable to the mean free path of the gas. However, such resolution would be prohibitively expensive if used to simulate the evolution of the entire vapor plume. Instead, the DSMC simulations are run in a series of nested staged domains (*Prem et al. 2015*). The DSMC molecules require the highest spatial and temporal resolution in the innermost domain; as particles leave the domain, their positions, velocities, and other properties are recorded and saved, to be later fed into the domain interface of the next domain out having a locally appropriate spatial and temporal resolution. In this way, staging, while complex, allows for our DSMC code to simulate the evolving plume more rapidly.

Nevertheless, staging itself is extremely time consuming, even though the initial evolution of the plume is only weakly sensitive to the properties (e.g., gravity) of the parent body itself. We therefore have chosen to skip the entire SOVA-DSMC hybridization process, and instead start our simulations by scaling the state of the evolving plume on the Moon from *Prem et al. (2015)* starting from 30 s post impact. To do this, we maintain the properties of the plume (mass, momentum, energy distributions and overall plume shape and details), and merely move it to lie at the surface of Mercury. Although Mercury has a greater radius (and therefore more gentle curvature) than the Moon, the plume is sufficiently small even 30 s post impact that the difference in curvature is sufficiently small that it affects a negligible number of simulated molecules. A small number of imported molecules located below the surface of Mercury due to this difference in curvature were removed from the simulation.

The errors associated with this approach are negligible. The difference in surface gravity between the Moon and Mercury is 2.1 $m/s^2$. After 30 s, this leads to a maximum error in the radial component of velocity of 63 m/s. This is small compared to the flow speeds generated by the 30 km/s impact; fewer than ~0.0025% of the molecules are moving slower than or comparable to this error in velocity magnitude (see figure 1). Additionally, although the rate of photodestruction of water is an order of magnitude faster at Mercury than the Moon, the density of the plume shields the overwhelming majority of particles from experiencing any difference in radiation environment during the first few minutes. Furthermore, the photodestruction timescale for an unshielded water molecule at 0.387 AU is ~3.6 hours (*Steckloff et al. 2022*), which is nevertheless 460 times longer than the elapsed 30 s of the restart files being used, and therefore ignoring the first 30 s of photodestruction leads to a negligible error in the number of particles that have been photodestroyed.

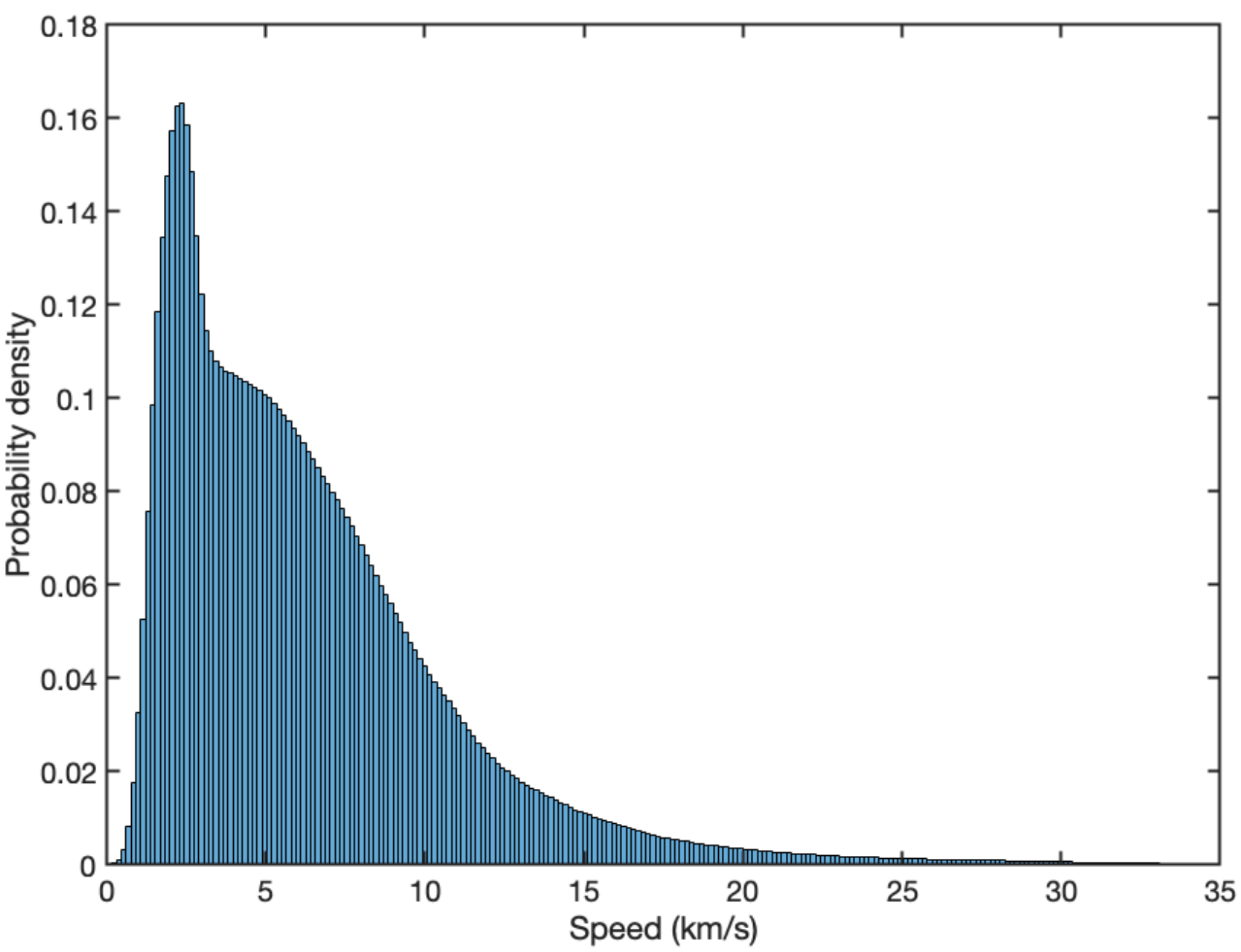

**Figure 1:** *Distribution of the simulated molecular speeds 30s post-impact.* The distribution of the molecular speeds at the start of our DSMC simulation, taken directly from the simulations of *Prem et al. (2015)* at 30 s post impact; these simulations model a comet impacting the north pole of the Moon at 30 km/s with an impact angle of 60 °. Although these molecular speeds are not normally distributed, they have a mean value of 6.3 km/s and standard deviation of 4.4 km/s with a long high-speed tail. In general, these molecular speeds are much greater than the 63 m/s error due to the differences between the Moon's and Mercury's gravity over the first 30 s of plume evolution.

*2.4 Cold Traps*

PLANET DSMC can incorporate lists of cold traps, modeled as circular projections on the surface of Mercury, with a trapping efficiency (probability of a molecule being cold-trapped upon impact) that can be varied from 0% to 100%. Cold traps have been observed extensively on Mercury, with both radar-bright materials and high-albedo surface materials indicative of water ice that correlates with low, permanently shadowed topography as measured by the MESSENGER spacecraft (*Paige et al. 2013; Chabot et al. 2014; Deutsch et al. 2016; Chabot et al. 2018*). We model these large, discrete cold traps as 100% efficient at trapping water molecules that impact their surfaces.

In addition, volatiles can also collect in unresolved, "diffuse" cold traps, which can be either small craters that are unresolved in spacecraft observations (*Rubanenko et al. 2018*) and/or the result of shadowing due to surface roughness (*Prem et al. 2018*). These diffuse cold traps appear to cover vast regions of Mercury's surface at high latitudes, albeit with low trapping efficiencies overall across these regions (*Deutsch et al. 2016; Chabot et al. 2018*). We model these diffuse cold traps as large, circular regions centered concentrically on Mercury's respective poles, with the probability that an incident water molecule happens to land in a micro cold trap (rather than a warm, illuminated element of the surface) of at most a few percent. In total, our model includes a total of forty discrete cold traps and five large, diffuse cold traps between the northern and southern polar regions on Mercury, to account for some of the latitude dependence of the micro cold traps comprising these diffuse cold trap bands (*Prem et al. 2019b*). See Tables A1 and A2 for a list of these cold traps, their locations, size, and trapping efficiency.

We can add up the areas of these cold traps to compute the total cold trapping area. To account for the non-planar geometry of Mercury's surface, we treat each of these cold traps as a spherical cap (or spherical "dome") that has a radius along the curved surface of Mercury; this is only significant for the large diffuse cold traps, for which the curvature of Mercury is significant. We then multiply their total area by their trapping efficiency, to compute the effective area of these cold traps. This results in a total of 21,887 km$^2$ of cold traps in Mercury's northern hemisphere and 34,283 km$^2$ in the southern hemisphere. We compare these totals with Feoktistova et al. (*2021*), who modeled cold traps totaling 23,300 km$^2$ in the north and 45,500 km$^2$ in the south. This results in a small difference of ~6% in the areas of the north polar impacts, but a much larger difference of ~25% in the south polar cold trap areas. These differences are most likely due to the vastly different number of cold traps included in our respective models. Whereas we have 23 north polar and 22 south polar cold traps, *Feoktistova et al (2021*) included 84 and 108 cold traps in the north and south polar regions, respectively.

## III. Results

We find that there are several distinct phases in the development of the impact plume, each defined by unique exosphere conditions and structures, beginning with the early plume phase, which describes the earliest phase of the plume evolution, lasting the first few hours from impact and through the initial expansion of the plume out to Mercury's Hill sphere. This is followed by the reentry phase, which describes the fallback of the gravitationally bound components of the plume, and the resulting formation of shock structures and frost across and above the surface; this phase lasts ~20 hours. As the plume settles out, the exosphere enters a Quasi-Steady phase in which the atmosphere enters a stable, steadily declining state that lasts for a few rotations of Mercury. Finally, the plume enters its late phase, in which the last lingering molecules bounce around the surface, and ultimately reach their final fate. Here, we describe our results in the context of these separate phases.

### *3.1 The Early Plume*

Early on, the plume retains its initial ejecta structure, with a significant enhancement of material traveling down-range due to the non-vertical, 60° impact angle. The plume is still densest near the point of impact, with densities as high as ~10$^{23}$ water molecules/m$^3$ and mean free paths as small as ~10 µm (see figure 2). At these high densities, these inner regions of the plume are still in the continuum regime, rather than the regime of rarefied flow.

To show this, we compute the Knudsen number of the plume. The Knudsen number compares the mean free path of the molecules with a characteristic length scale of the situation, to determine the importance of collisions on the evolution of the flow field over that characteristic length. Flow is considered to be in the highly collisional, continuum regime for Knudsen numbers below ~0.01 and in free molecular flow for Knudsen numbers greater than ~10. In either of these regimes, DSMC may be unnecessary for our purpose – because in continuum flow the flow properties vary sufficiently slowly compared to the characteristic density gradient length scales such that collisions dominate and maintain near-thermodynamic equilibrium in the flow field, and collisions in free molecular flow are sufficiently rare that molecules behave as isolated ballistic masses. It is in the transitional, Knudsen flow regime where collisions can significantly impact the flow and evolution of the vapor plume over length scales comparable to the mean free path, that techniques such as DSMC are required to solve the Boltzmann equation and extract the gas dynamics.

For our purposes, we compare the mean free path (λ) with the characteristic length scale over which the density gradient changes ($L_{gradient}$)

$$\lambda = \frac{1}{\sqrt{2}\rho\sigma} \tag{2}$$

$$L_{gradient} = \frac{\rho}{|\nabla\rho|} \tag{3}$$

where σ is the effective collisional cross-sectional area of the molecule, ρ is the number density and ∇ is the gradient operator. Note that the gradient and density, and thus the gradient lengthscale varies across the plume. Thus, the local gradient-based Knudsen number (*Kn*) of the plume is

$$Kn = \frac{\lambda}{L_{gradient}} = \frac{\lambda\,\nabla\rho}{\rho} \tag{4}$$

$$= \frac{\nabla\rho}{\sqrt{2}\rho^2\sigma}. \tag{5}$$

The early plume is almost exclusively in the continuum flow regime, with only isolated rarefied regions on the margins of the plume (figure 2). Therefore, while we use DSMC to evolve the plume, we could have used a continuum solver instead, except at the rapidly changing plume periphery. Our cell sizes in DSMC are also much larger than the local mean free path in this region as the cost of simulating the additional collisions could be prohibitive. However, we use collision limiting (*Stewart et al. 2011*) to keep that cost within bounds, since we know *a priori* that those dense gas cells will be nearly in local thermodynamic equilibrium.

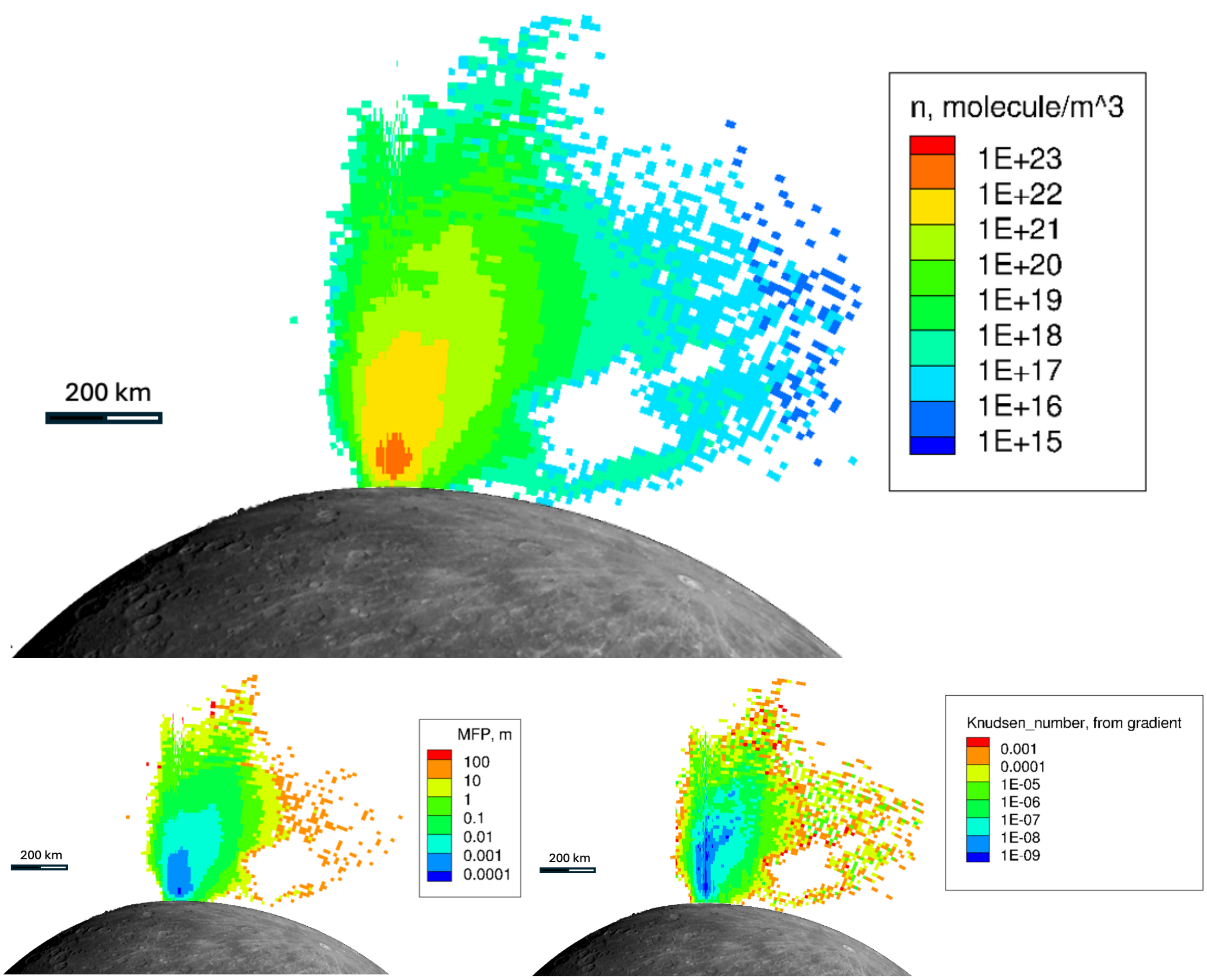

**Figure 2:** *Initial structure of the Impact Plume 60 s post impact.* The impactor strikes from the left, at a 60° angle from horizontal, with the sun located to the right.There is a visible asymmetry in the plume due to the direction of impact; the "hole" on the right side of the plume is a bubble in this 2D slice of a fully 3D plume due to the removal of water vapor-rock mixtures from the simulation during the sampling from SOVA into DSMC codes (*Prem et al. 2015*). *(Top)* the plume contains a region of high density (~$10^{22}$ molecules/m$^3$) just above the point of impact, which rapidly drops toward the edges of the plume. *(Bottom, left)* The high plume densities correspond to short mean free paths for the vapor molecules. *(Bottom, right)* Comparing this mean free path with the length scale of the density gradient reveals very small Knudsen numbers throughout; the entire plume (except for the very outermost edges) is clearly within the continuum flow regime at this time, rather than the slip-flow, Knudsen, or Free Molecular regimes.

In the earliest stages of the plume development, the high density of the vapor leads to continuum fluid dynamics. Since fluids flow from high to low pressure regions, the vapor expands away from the point of impact. The resulting gasdynamic acceleration significantly impacts the amount of vapor retained by the system. At our earliest time step (30 s post impact), ~54.5% of the molecules have sufficient kinetic energy to escape the Mercury system (if unimpeded), while ~45.5% of the molecules are gravitationally bound to the system. However, only 30 s later (1 min post impact), gas dynamics will have accelerated molecules such that 63% of the molecules are moving fast enough to escape the system, leaving only ~37% of the molecules gravitationally bound to Mercury. As the plume expands, its pressure gradients and density drop, eliminating the ability of the pressure field to accelerate significant numbers of gas molecules; the molecular speed distribution reaches an asymptote by 10 mins post impact such that ~65% of the molecules are moving fast enough to escape and ~35% remaining gravitationally bound. Thus, gas dynamic expansion after 30 s has increased the ballistically escaping fraction of gas molecules by nearly ~20%.

Solar EUV photons contain sufficient energy to photolytically destroy water molecules; as the plume evolves and expands, it exposes an ever larger cross sectional area to the Sun. This leads to a steadily increasing rate of photolytic destruction of water molecules in the expanding plume. Nevertheless, the plume is optically thick at the absorbing frequencies, with sunward water molecules absorbing these photons and shielding the water molecules in their shadows from photodestruction. Thus, sunlight begins to destroy the plume from the sunward surface. We note that the PLANET DSMC code does not presently track the molecular products of this photodestruction (H and OH in this case); any molecule that is photodestroyed is removed from the simulation, and its properties recorded.

This approach inherently overcounts the net effects of photodestruction, as it assumes no recombination of daughter products occurs to repopulate the plume with $H_2O$ and does not account for the UV opacity of those daughters. Nevertheless, the rarefied nature of the plume, presence of a significant interplanetary and planetary magnetic field (e.g., *Ness et al. 1976*), and high recoil speeds following a photodestruction event of ~1 km/s and ~17 km/s for OH and H respectively (*Crovisier, 1989*) imply that any daughter species are generally rapidly removed from the system, and that that recombination is a relatively rare event. Ultimately, this is simply an approximation to the usual result that the excess energy of UV photodissociation of $H_2O$ often leads to both daughter particles having enough energy (*Crovisier, 1989*) to escape. For similar reasons, we neglect H + OH recombination reactions as being highly improbable in the thin plume gas. PLANET can in fact handle multiple species and recombinative chemistry; perhaps future work could examine these effects.

The timescale of unshielded photodestruction of $H_2O$ molecules at 0.38 AU is ~3.6 hrs, which is comparable to the timescale of this initial phase of early plume evolution (the time over which fast-moving molecules in the plume reach Mercury's Hill Radius (175,000 km from Mercury at 0.38 AU) or begin to fall back to the surface. As a result, a large fraction of the ~65% of the molecules that are moving sufficiently fast to leave the Mercury system are photodestroyed in the rapidly thinning plume prior to reaching the Hill sphere. PLANET DSMC counts such molecules as photodestroyed, rather than ballistically escaped; comparing the fraction of molecules that PLANET DSMC records as ballistically escaped (13.6% of launched $H_2O$ molecules would make it to the Hill sphere intact) with the fraction (~65%) that have sufficient energy to escape reveals that 79% of the escaping molecules are photodestroyed prior to reaching $R_{Hill}$ and leaving the system. Ultimately, we find that ~68% of the initial water molecules are lost from the Mercury system through photodestruction or ballistic escape during the first 24 hours post impact.

This is a significant departure from what would be expected without self-shielding. Given a photodestruction timescale of ~3.6 hrs (which quantifies the time over which the number of exposed molecules would drop by a factor of 1/*e*), a water molecule would only have a ~0.13% chance of surviving 24 hours of exposure to sunlight, and a ~99.9% chance of being photodestroyed. Granted, many molecules land on the surface within the first 24 hours (adsorbed molecules do not photo-dissociate in our simulations). However, self-shielding of gas within the rising/falling plume qualitatively increases the survival probability considerably.

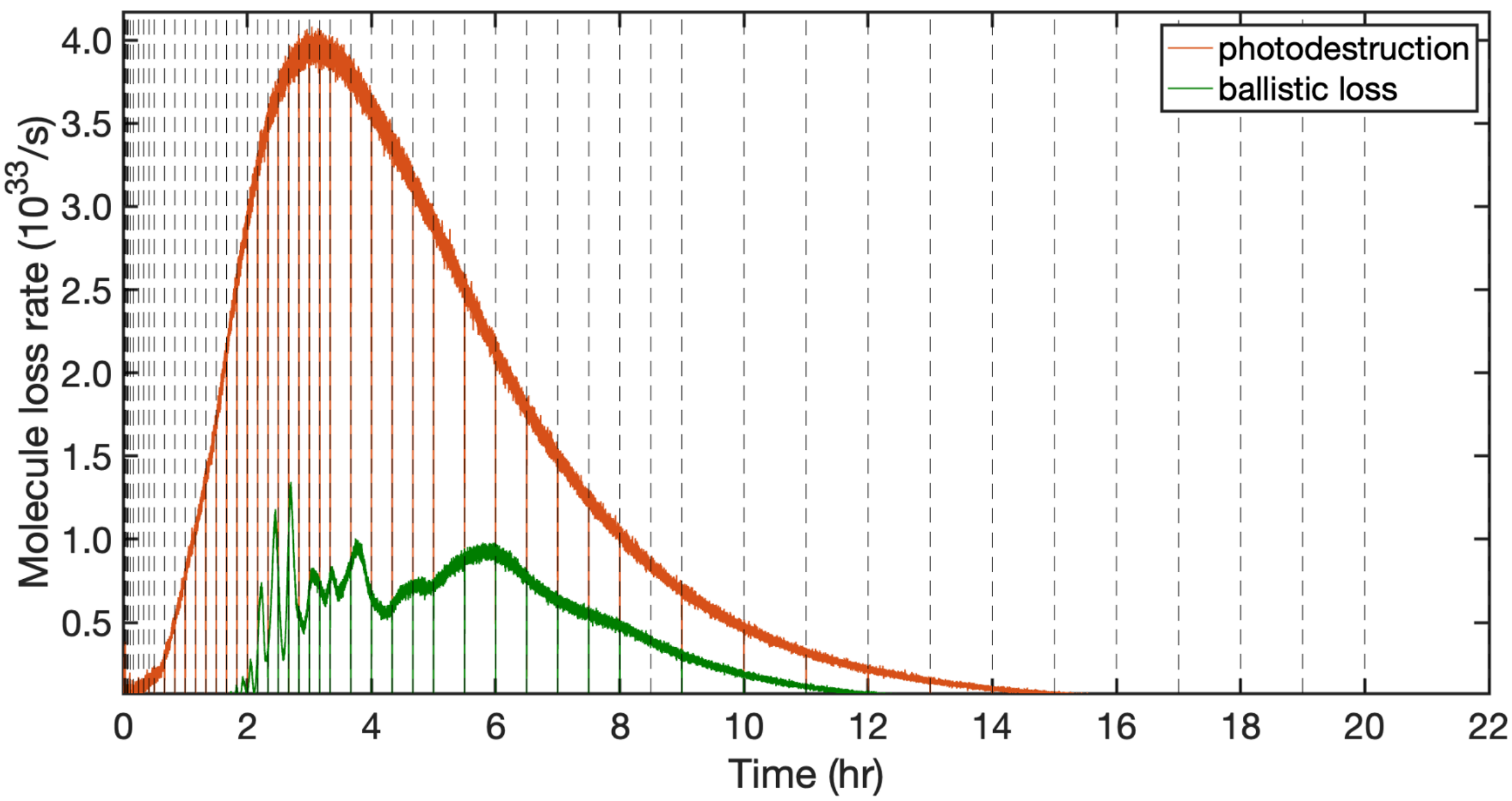

**Figure 3:** *Ballistic escape and Photodestruction Rates in the first 22 hrs Post-Impact.* As the plume expands post-impact, the plume cross section exposed to the Sun expands, dramatically increasing the photodestruction rate, before part of the plume begins to collapse under Mercury's gravity and drops back down. As the plume expands, particles travel toward, and eventually reach Mercury's Hill Sphere, where they are lost from the simulation and recorded as a ballistically escaping molecules; due to this accounting scheme, the ballistic loss and photodestruction rate profiles reach maxima around the same time (peaking ~2-6 hrs post impact). The ballistic loss profile is more erratic due to density asymmetries or lumps in the plume as they cross Mercury's Hill Sphere. Vertical dashed lines denote simulation restarts; at each restart, both the ballistic escape and photodestruction rates drop to zero for a single timestep, before being recalculated.

Eventually, the vapor plume stops expanding, and begins to collapse back on itself, as molecules are slowed by Mercury's gravity, and begin to fall back toward the surface. This happens quickly; the rate of photodestruction is proportional to its opacity-weighted cross sectional area with the Sun, and peaks approximately 4 to 6 hrs post impact (see figure 3). As the remaining water molecules fall back toward the surface of Mercury, they eventually collide with the surface/growing atmosphere, triggering the start of the reentry phase of the plume evolution. As the last molecules cross Mercury's Hill sphere approximately 20 hrs post impact, the ballistic escape rate of intact water molecules drops to effectively zero for the remainder of the simulation (further discussion on this topic in the Quasi-Steady State section 3.3).

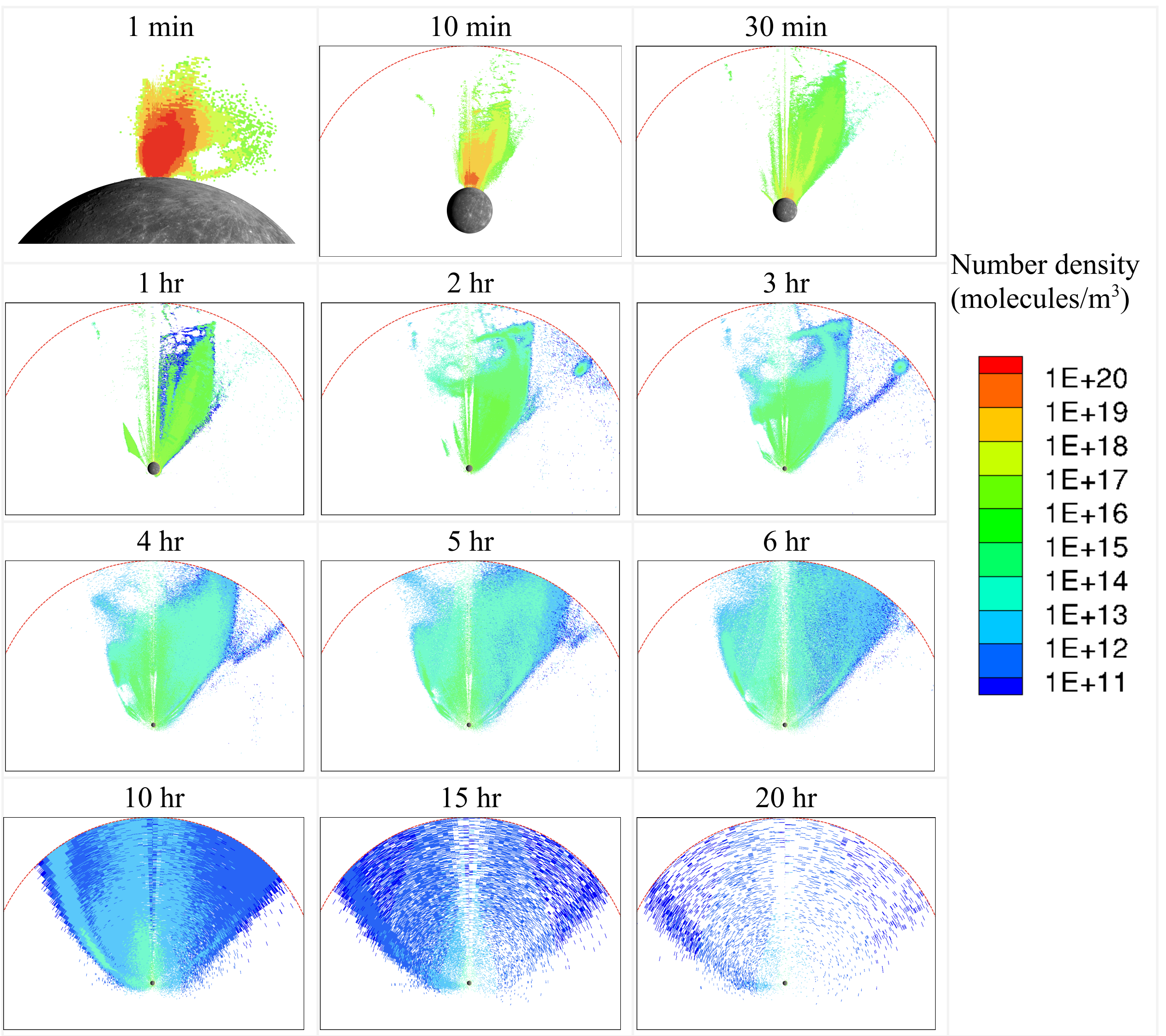

**Figure 4:** *Density structure of the evolving early plume.* The vapor plume that results from the impact evolves rapidly as the plume expands from the point of impact, producing an evolving structure with time seen in these slices. Mercury is included for scale in each frame; red dotted line denotes the edge of the output grid of our numerical mesh; the full numerical domain is much larger and stretches to Mercury's Hill sphere. This evolution slows as the plume expands, with notable features in the plume moving little at later time steps in the early plume. Note that there are several numerical artifacts associated with insufficient grid or particle count in some small regions or changes in grid or time step during occasional restarts. This is particularly acute near the axis of the computational mesh; due to spherical coordinates being used, these cells have small volumes, and are therefore more likely to be "empty" which registers as a blank space in the data.

### *3.2 Reentry*

As the plume expands, molecules that have sufficient energy to escape Mercury continue to rise, while most molecules moving slower than escape speed eventually turn around and fall back toward the surface as they continue to photodissociate This creates a phase of the evolving plume defined by the collapse of the inner vapor plume as it falls back upon itself, causing the molecules to collide, compress, and heat up. This reentry phase is defined by the period over which the falling evolving gas plume contacts the ground and compresses other parts of the plume, and ends once the impact plume settles back onto the surface and the resulting atmospheric disturbances damp out (~20 hours post impact).

Slower moving molecules from the initial plume impact the surface almost immediately after impact, falling near the impact site. Over time, progressively faster-moving molecules impact the surface at ever greater distances from the impact site, forming an advancing curtain or cloud of water vapor that sweeps outward across the planet from the impact site. When this vapor front reaches Mercury's cold traps, the cold traps begin to fill with water molecules, As the plume spreads out from the impact point at the North Pole, the cold traps are initially populated from north to south, leading to a latitudinal dependence in the initial population of the cold traps.

The northern cold traps begin to populate immediately, with all discrete northern cold traps starting to receive water molecules within the first 200 s post impact. By contrast, the southern cold traps only start receiving water molecules 3500 - 4200 s post impact. This difference relates to the ballistic timescale associated with Mercury, the period of an orbit that just skims the surface of Mercury (assuming a perfect sphere) is 85.2 mins, or 42.6 mins (2560 s) to travel half that distance, and is the timescale over which the impact at the North pole begins to populate cold traps near the South Pole. In actuality, the molecules are following higher/slower ballistic trajectories that take a little longer to reach the south, explaining the slightly longer delay in populating the southern cold traps. Regardless, this initial process of populating Mercury's cold traps with water molecules has a latitudinal dependence as the plume sweeps across the surface from north to south.

This initial populating of the cold traps continues for as long as the initial plume continues to fall back onto the surface and deposit molecules in them, ending ~20 hrs post-impact, at which point the initial plume is observed to have dissipated (see figure 4), due to ballistic escape, cold trapping, surface adsorption, and photodissociation. Approximately ~1% of the initial water content of the comet nucleus is trapped in Mercury's various cold traps during this initial period.

*3.2.1 Reentry Shock*

As water vapor falls back onto the hot illuminated surface of Mercury, it momentarily sticks to the surface and thermally equilibrates, before being quickly reemitted (*Langmuir, 1916*), and beginning to hop ballistically across the surface. However, further fallback of water vapor causes the water vapor density on the illuminated hemisphere to increase until this collective $H_2O$ gas becomes a collisional atmosphere. Once collisional, the $H_2O$, which continues to fall back to the surface, begins to collide with, and compress the gas already on/near the surface, increasing its density and temperature. When sufficiently dense, these falling $H_2O$ molecules deposit their momentum and energy into this rising atmosphere, forming a shock layer at the top that is hotter than the rest of the dayside atmosphere (see figure 5).

This reentry shock layer serves as a stationary pusher plate, maintaining compression of the atmosphere. As a result, this atmosphere is ~250 km thick from surface to the top of the thermal shock layer (the thermal shock layer itself is only ~30 km thick) While this shock layer serves to inhibit the atmosphere from expanding radially from the surface, it allows the atmospheric layer to flow along the surface; aided by the oblique or non-vertical velocity components of the reentering molecules, the high-altitude wind within and just below the shock layer dynamically pushes the flow generally toward the South Pole. That is, the returning high speed gas molecules moving at several km/s are driving these high-altitude winds, which flow toward the south pole at ~300 m/s. However, the atmosphere near the ground, well below the shock layer, is isolated from the motion of the reentering molecules, and instead experiences pressure-driven flow toward the terminator. Beyond the terminators, the pressure drops to near vacuum as molecules adsorb onto the surface. This lower altitude flow is correspondingly slow near the subsolar point, with winds of up to a few 10s m/s; these winds pick up away from the subsolar point, reaching ~100 m/s speeds near the terminator.

Thus, the surface atmosphere can experience two distinct flow layers driven by independent mechanisms (see figure 5). We note that the direction of the low altitude flow is driven by Mercury's surface temperature and is toward the night side; the high altitude shock layer wind direction depends on the location and angle of the initial impact. The gas density within the shock layer and reentering plume is sufficiently great to absorb the vast majority of incoming solar photons with sufficient energy to photodestroy water molecules. As a result, the atmosphere below the shock layer is largely shielded from photodestruction.

The reentering molecules are kinetically cool (high Mach number) but they heat up as they collide with other gas molecules in the atmosphere, generating an extremely hot shock layer with temperatures on the order of ~1000 K; below, the dayside atmosphere is at a uniform ~500 K. The density of this surface atmosphere is gravitationally controlled with a density at the surface on the order of ~$10^{18}$ molecules/m$^3$, dropping to ~$10^{17}$ molecules/m$^3$ by altitude ~100 km.

From these quantities, we can emperically compute the atmospheric scale height (*H*), which defines the vertical distance over which the density of the atmosphere drips by a factor of $1/e$. This corresponds to a scale height of ~43 km. We compare this to the expected scale height, which can be computed with the equation

$$H = \frac{k_B T}{\mu g} \quad (6)$$

Where $k_B$ is Boltzmann's constant, $T$ is translational temperature, μ is the molecular mass, and $g$ is the acceleration due to gravity. We find that the empirically measured scale height is roughly equal to the expected scale height from the translational temperatures throughout the dayside atmosphere of ~35-50 km.

Although the water vapor in these simulations experiences a wide range of pressures and translational temperatures, the pressures are nevertheless everywhere too low for condensed phases to form (in the atmosphere/exosphere, outside of the cold surface locations). Thus, we expect volatile migration to the terminator on Mercury. This is fundamentally different from similar simulations conducted for the Moon; *Prem et al. (2015)* found that the analogous surface atmosphere on the Moon has comparable number densities (~$10^{18}$ molecules/m$^3$) but is sufficiently cool for the local vapor pressure to exceed the equilibrium vapor pressure of pure water. In such situations, the water vapor could condense, leading to potential ice crystal formation and falling "sleet" on the Moon *(Prem et al. 2015)*. This difference is likely due to two factors. First, *Prem et al.* (*2015*) did not model radiative heat transfer within the vapor plume in their DSMC simulations; radiative heating enables hot parts of the plume and the surface to emit thermal photons that can then be absorbed by the vapor within the plume, heating up cooler parts of the plume. Second, the extreme surface temperatures of Mercury prevent condensed phases from forming. Upon colliding with the surface, the vapor molecules thermalize, matching Mercury's ~400 K mean surface temperature. Also, the surface gravity of Mercury is much greater so returning gas molecules are much faster than at the Moon and the sub-shock temperature thus much greater. Together, these conditions preclude any solid or liquid water formation above the surface.

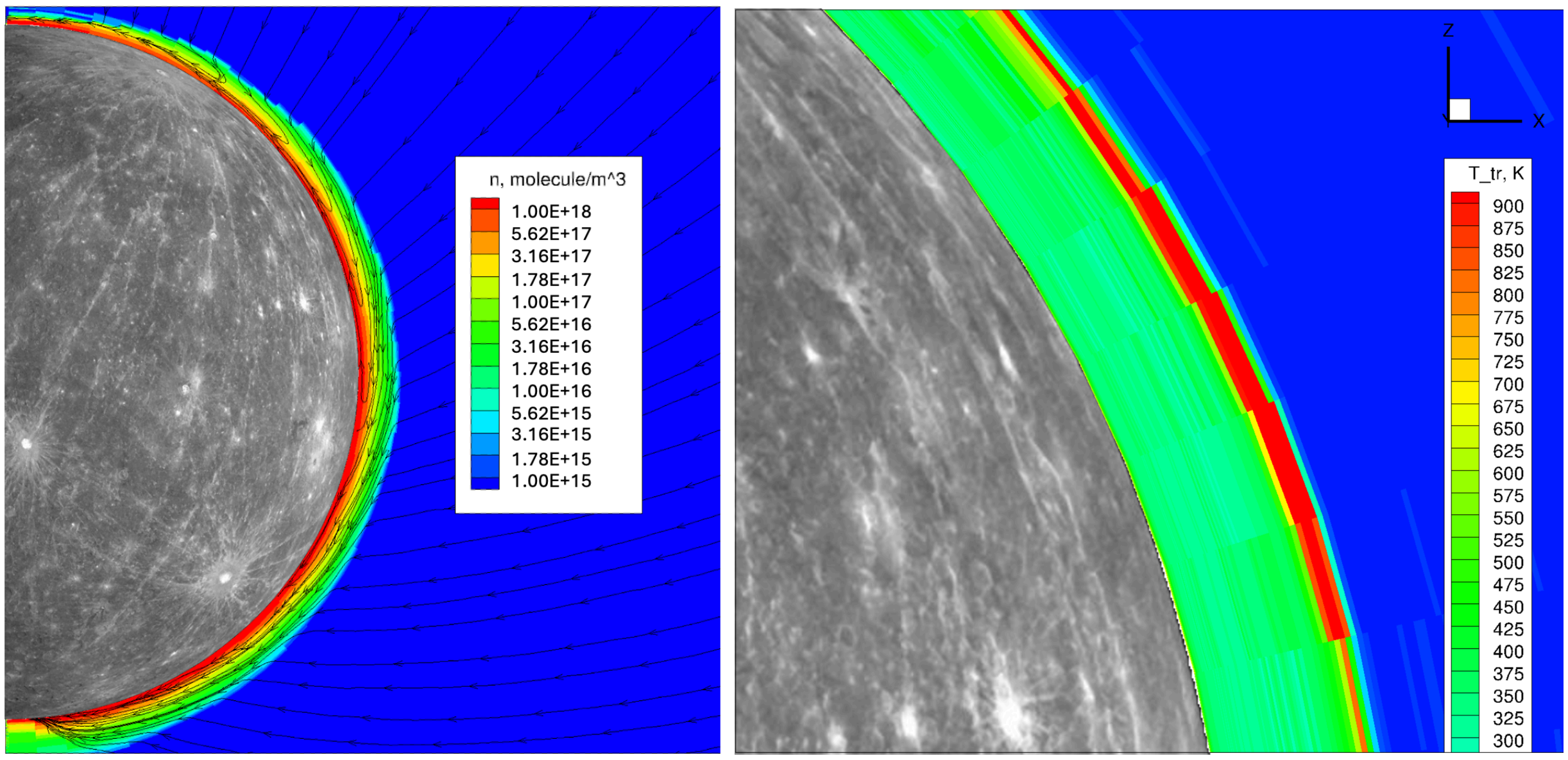

**Figure 5:** *Reentry shock density, translational temperature, and radiative heating rate 6hr post-impact.* As material falls back and collides with other water vapor, it slows down and heats up, forming a high-altitude reentry shock layer that compresses the atmosphere. (*Left*) the atmosphere forms a typical structure with peak density below, and maintained by dynamic forces. This produces a "typical" atmospheric pressure profile that decreases exponentially with altitude until the shock layer, above which pressure drops precipitously. (*Right*) the translational temperatures are extremely high in the shock layer, on the order of ~1000 K, but are nearly isothermal in the main body of the atmosphere.

*3.2.2 Antipodal Shock*

Following impact, the vapor plume initially rarefies as it spreads approximately radially out from the impact site. However, Mercury's gravity causes some of the ballistic trajectories to curve, eventually converging at the antipode of the impact site. This antipodal convergence creates a columnar shock over Mercury's south pole (see figure 6). We note that the shock has an anti-sunward curve with increasing distance over the pole; this is due to the non-axial nature of the initial impactor, which struck the surface at an angle of 60° with a ground track toward the Sun. Such an impact releases significantly more material downrange (sunward in this case), which dominates the mass balance at the point of convergence and subsequently "pushes" the shock wave anti-sunward.

The density of this antipodal columnar shock structure is ~$2 \times 10^{17}$ molecules/m$^3$ (fig. 6), roughly an order of magnitude greater than the reentry shock wave density seen in Fig. 5, corresponding to a pressure of ~1 mPa (~10 nbar). The temperatures in the antipodal shock of ~600-800 K are much cooler than the surface shock layer, which has temperatures around ~900 K. We note one key feature of the antipodal shock contributing to its relatively low temperature is consideration that a large fraction of the gas contributing to the converging flow first traveled over the night side of Mercury, where it could cool via infrared emission from rotational-vibrational (rovibrational) transitions and not be rewarmed by sunlight.

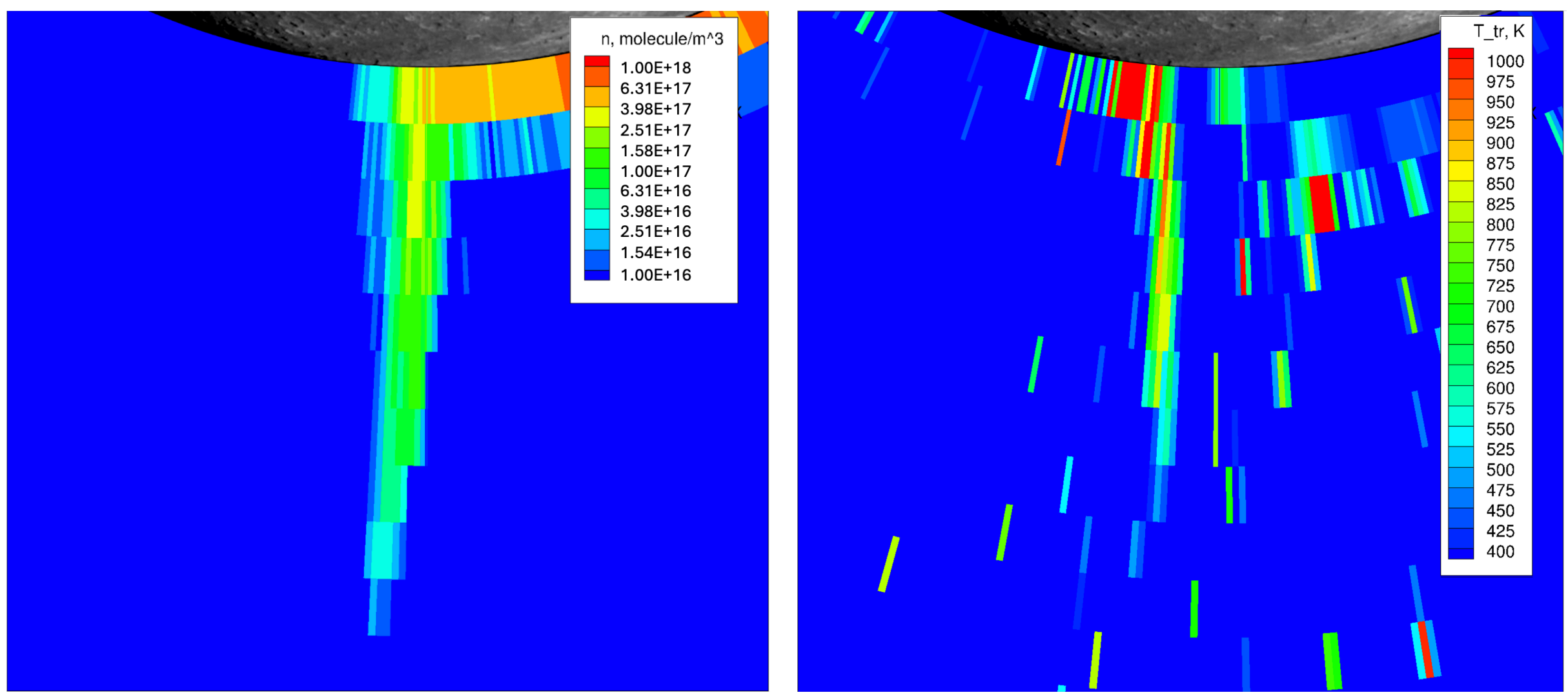

**Figure 6:** *Antipodal shock density and translational temperature 6hr post-impact.* The expanding vapor plume collides with itself at the antipode of the impact site, where the ballistic trajectories of the gas molecules converge. The collisions heat the gas and slow the molecules, resulting in a columnar shock above the antipode with *(left)* densities on the order of ~$2\times10^{17}$ molecules/m$^3$ and *(right)* temperatures of ~800 K, corresponding to a characteristic pressure of ~10 nbar.

*3.2.3 Frost Distribution and Initial Population of the Cold Traps*

Once material ceases to fall back, and the reentry shock features dissipate, we can investigate the initial distribution of impact-delivered surface frost adsorbed onto Mercury's surface. As the simulated impact was moving along a ground-projected path pointing toward the sub-solar point, particles ejected uprange necessarily travel along the night side of the planet. Any molecules that strike the night side surface were immediately stick (i.e., are adsorbed) onto the surface due to the cold (100K) temperatures present on Mercury's nighttime hemisphere. Conversely, water molecules that travel down range in our simulation will often strike the hot illuminated hemisphere of the planet, where high surface temperatures preclude such adsorption onto the surface. The exception to this would be the cold traps/PSRs on the illuminated hemisphere and the ribbon of the surface just on the sunward side of the terminator, where surface temperatures are sufficiently cool to allow water molecules to adsorb. Instead, these molecules are effectively re-emitted immediately, and will continue to hop and collide with the surface and each other until they are either photodestroyed, land in a cold trap, or migrate to the cold unilluminated surface and are adsorbed. Since the ballistic hop distance of these molecules is small compared to the size of Mercury (*Steckloff et al. 2022*), those molecules headed toward the night side generally adsorb near the terminator, leading to a significant enhancement in the initial frost concentrations near the terminator at the time of impact (see figure 7).

The initial frost distribution also shows a significant dependence on latitude. Unsurprisingly, frost concentrations near the point of impact are among the highest in our simulation, with the frost concentration becoming smaller with increasing distance from the impact site. This trend continues out to the equator of Mercury, beyond which (i.e., south of ~40° S) the frost density increases, reaching a second peak density near the South Pole. This increase beyond the equator is due to the plume reconverging as it approaches the South Pole; effectively gravitational focusing causes streamlines to converge toward the antipode. Notably, the minimum frost density occurs not at the equator, as would be expected by purely geometric concerns, but rather at ~20° S. This is due to ejecta thinning with increased distance, which pushes this minimum further south. See supplementary figure S1 for a spherical geometry-corrected plot, showing how the frost distribution is generally consistent with ejecta thickness decreasing with distance from the impact site.

South of ~70° S there is a significant enhancement in the amount of frost deposited, relative to what would be expected due to antipodal focusing or distance from impact site (see figure S1). This Southern enhancement is due to the lack of a collisional atmosphere forming on the unilluminated side of Mercury. As the atmosphere and fall back material on the illuminated hemisphere reaches the South Pole, there is significantly less material arriving at the South Pole from the unilluminated hemisphere as any material that struck the cold, unilluminated surface adsorbed onto it. Thus, there is significantly more mass/momentum reaching the antipode from the illuminated hemisphere than from the unilluminated hemisphere; this pushes the point of convergence away from the south pole and toward the unilluminated hemisphere, leading to an enhanced frost deposit between ~70-90° S.

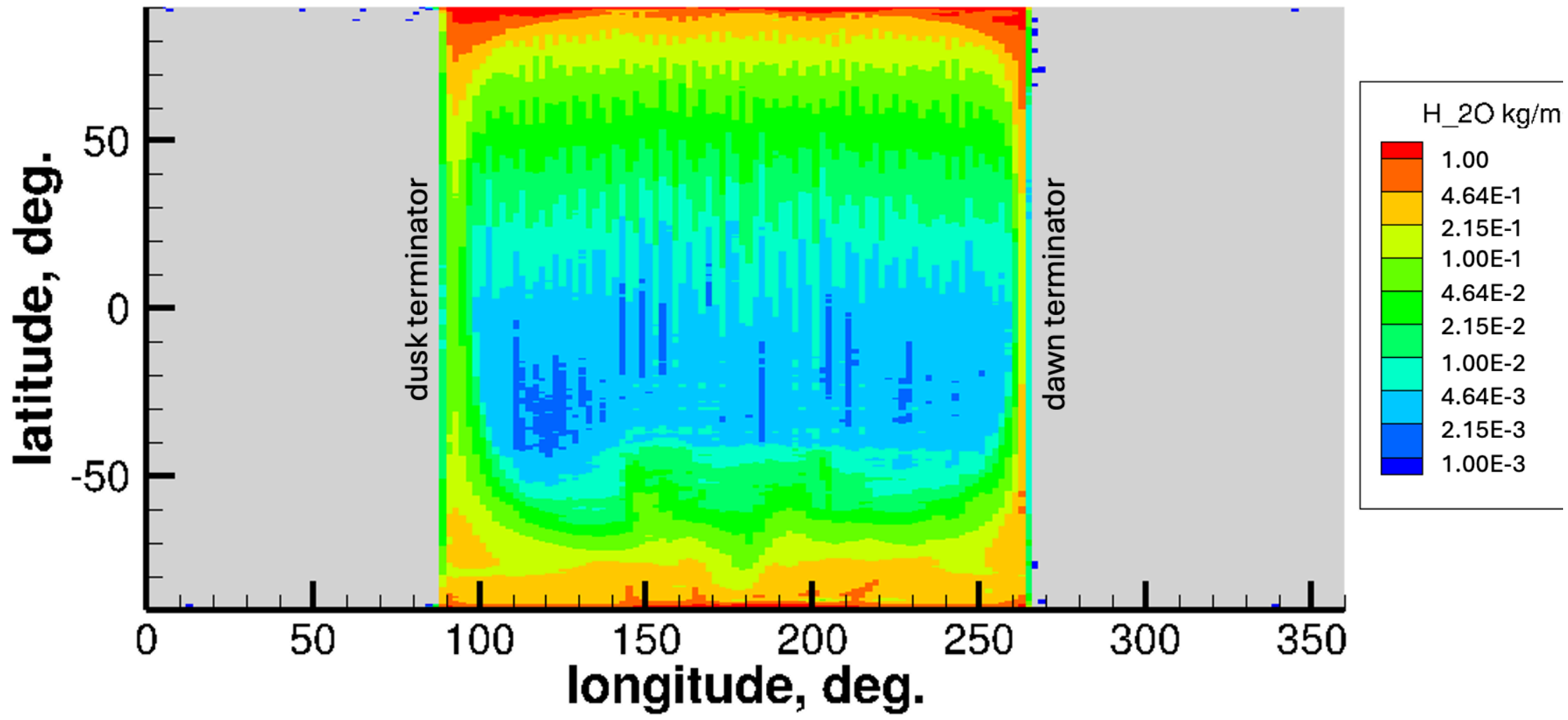

North Pole South Pole

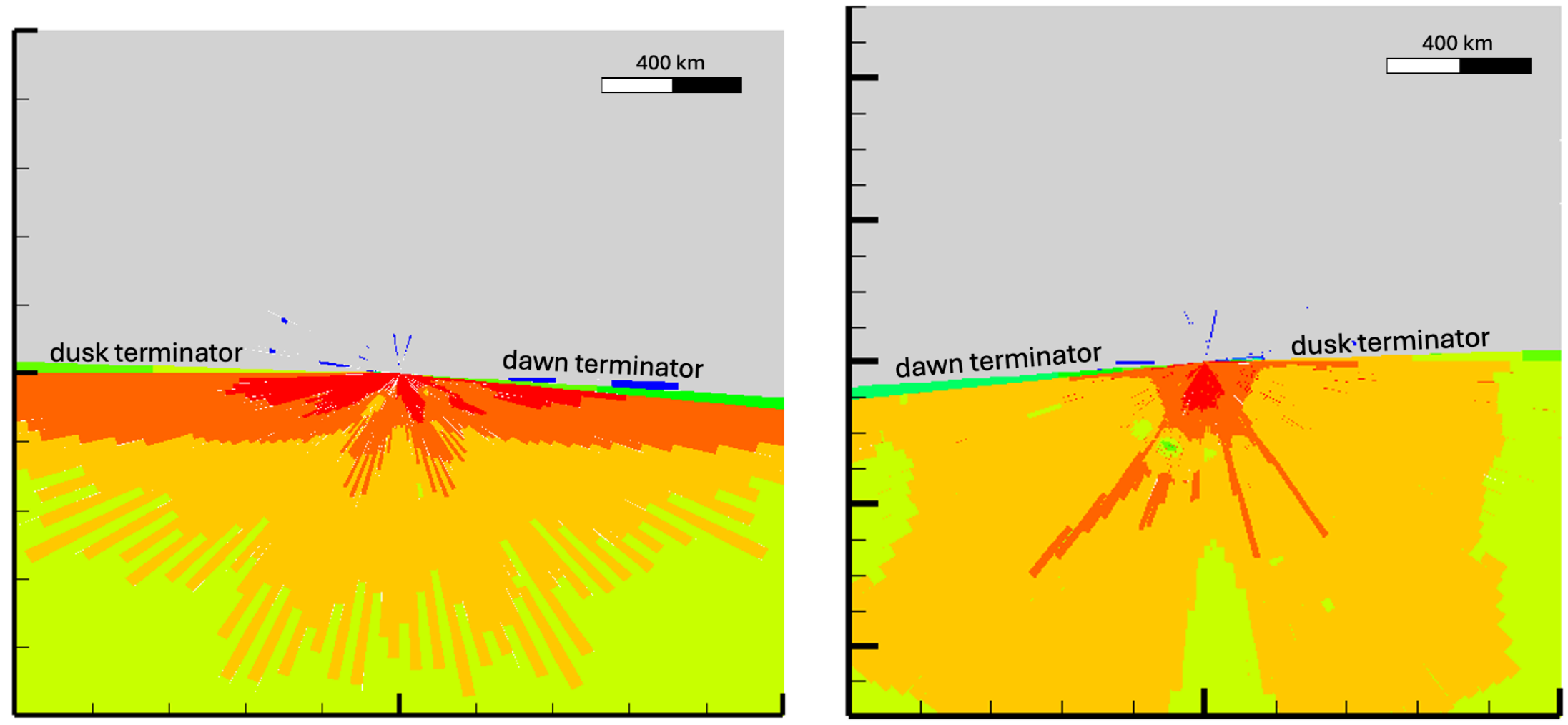

**Figure 7:** *Initial frost distribution 36 hr post-impact.* The initial frost is nearly fully adsorbed onto the surface of Mercury following the end of the reentry phase. *(Top)* This figure is centered on the unilluminated hemisphere (subsolar longitude is 0/360°) to keep the adsorbed region contiguous. Enhanced deposition at the terminator is observed, consistent with gas within this surface atmosphere flowing and adsorbing onto the surface. *(Lower left)* the north polar region near the point of impact has the highest frost densities, as plume density decreases with distance from the impact site. *(Lower right)* antipodal convergence produces enhanced frost density around the antipode (south pole). The peak south polar frost density is offset anti-sunward from the south pole, perhaps due to the anti-sunward displacement of the south polar columnar shock. Due to warm temperatures, no water ice is adsorbed on the illuminated hemisphere outside of the cold traps.

There are also numerical artifacts visible in this frost distribution, as linear north-south striations of increased frost density that produce non-physical enhancements along some lines of longitude. Nevertheless, these artifacts disappear after rotating into the dawn and smoothing out in space after desorbing and moving through the atmosphere, causing a short-lived, negligible pulsing in atmospheric density that rapidly damps out.

The reentering plume populates Mercury's cold traps asymmetrically. Because the simulated impact struck Mercury's north pole, the reentering plume sweeps the surface from north to south. As a result, Mercury's cold traps are populated latitudinally, from north to south, with the southern cold traps beginning to fill after a local ballistic timescale. Once the ejecta begins to arrive in the south, deposition into the southern cold traps dominates deposition into the northern cold traps for the remainder of the reentry phase. By the end of the reentry phase, 1% of the initial water molecules have become trapped in the cold traps; ~0.6% and ~0.4% in the northern and southern cold traps respectively.

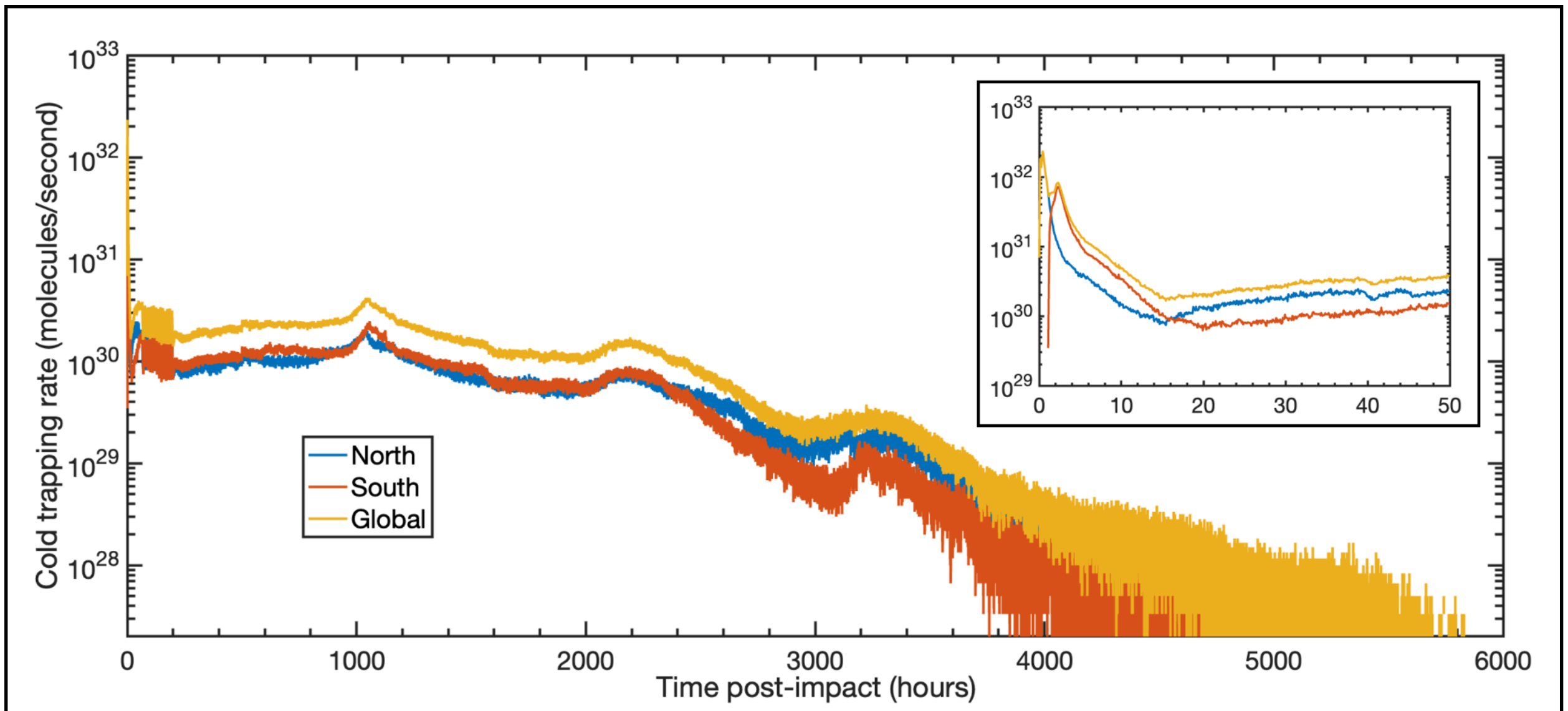

**Figure 8:** *Cold trapping rate as a function of time.* We plot the number of molecules per second falling into the northern cold traps, southern cold traps, and the sum of the two together. To reduce noise, data are averaged over 5 min increments. Although the northern cold traps begin to populate immediately, the southern cold traps only begin to populate once the ejecta curtain arrives in the south polar region approximately one hour post impact (a delay comparable to the ballistic timescale or half an orbital period). Thereafter, deposition into the southern cold traps dominates until the end of the reentry phase ~15-20 hrs post impact. After the collapse of the initial plume, the system enters a quasi-steady state, where the cold trapping rate remains approximately constant for a full rotation (hence the small peak at ~1100 hrs and another a day later at ~2200 hrs, etc), before steadily declining. The oscillations at ~50-200 hrs are a numerical artifact.

*3.3 Quasi-Steady State*

Eventually, the remaining plume settles back onto the surface, ending the direct effects of the impact on the structure of the atmo/exosphere; instead, the atmo/exosphere structure is determined by the slow rotation of adsorbed frost into daylight and resulting dynamics upon desorption. As this latter process varies very slowly relative to the timescales of atmospheric/exospheric dynamics, the resulting structure is stable and nearly steady. Thus, the settling of the initial impact plume onto the surface ends the initial fully 3D, momentum-driven unsteady flow phase, and marks the transition into quasi-steady, more diffusive flow. The transition from unsteady plume to quasi-steady exosphere occurs approximately ~20 hours post impact (roughly the rise/fall time to the Hill radius) and can be seen clearly in the inset to Fig. 8. At this transition point, the ballistic loss of molecules (molecules with sufficient kinetic energy to reach Mercury's Hill sphere and escape the system) drops to virtually zero, as thermal speeds at Mercury's surface are too slow for any significant fraction of molecules to reach escape speed. Furthermore, the rate of photodestruction of molecules declines to an approximately constant value, as the cross sectional area of the plume/exosphere drops to a constant value when the plume has fully settled back onto the surface. The quasi-steady state ends when the Dawn

Atmospheric Enhancement (see section 3.3.1) becomes optically thin, and photodestruction decreases to a final, terminal low level rate; this occurs a little after one complete rotation of Mercury.

*3.3.1 Dawn Atmospheric Enhancement*

As Mercury rotates, the adsorbed frosts on the night side slowly rotate into daylight, and warm up. Shortly after dawn, these surface frosts become sufficiently warm to desorb or sublime molecules from the surface. This forms a linear region running north and south at local dawn where peak desorption from the surface occurs. In turn, this forms a coincident region of enhanced exospheric/atmospheric density near the dawn terminator, as desorbed molecules leave the surface before scattering away. This Dawn Atmospheric Enhancement (DAE) phenomenon was first observed in Apollo 17 Lunar Mass Spectrometer data (*Hoffman and Hodges 1975*), and has subsequently been observed in comparable simulations of impact-induced atmospheres on the Moon (*Stewart et al. 2011; Prem et al. 2015*) and Io (*Walker et al. 2010; 2012*). Since DAEs are the natural result of a condensable gas on a body (*Hoffman and Hodges 1975*), they are expected to occur across a wide range of airless solar system bodies (*Steckloff et al. 2022*). However Mercury's greater surface gravity results in a narrower DAE than would be expected on the Moon or elsewhere in the solar system (*Steckloff et al. 2022*) outside of perturbing processes such as DAE-coincident shock layers that form following Io's egress from jovian eclipses (*Walker et al. 2012*).

The mean distance that a desorbed molecule "hops" is both a function of surface gravity and temperature. The hop distance decreases rapidly with increasing gravity, as molecules travel for less time (shorter distances) before encountering the ground. Nevertheless, the surface temperature at the DAE is approximately constant regardless of distance from the Sun, as adsorbed molecules will typically desorb as they rotate past the dawn terminator once the surface reaches a temperature warm enough for the residence time (*Frenkel 1924*) to become comparable to the synodic rotation period (versus orders of magnitude longer than the rotation period on the unilluminated side). Thus, the thermal velocities of the desorbing molecules, which are proportional to $\sim \sqrt{k_B T_{surface}}$, are approximately constant, resulting in a ballistic hop distance that is effectively controlled by surface gravity. If we consider the half of the desorbed molecules at the DAE that hop toward the night side, we can assume that each such molecule immediately sticks upon landing, as these surface temperatures are cooler than those at the DAE. Thus, the distances that these molecules cover before sticking is effectively one ballistic hop.

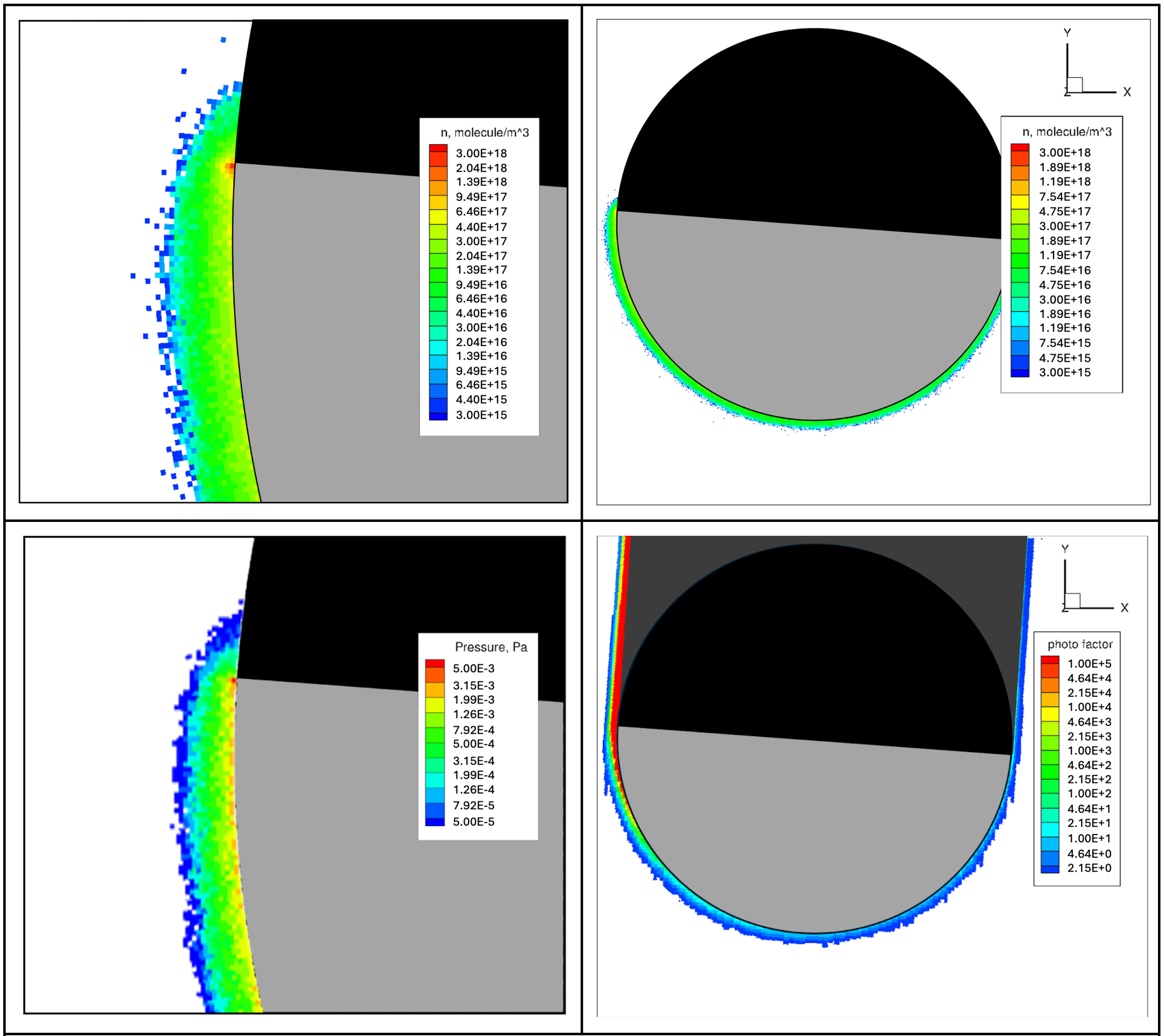

**Figure 9:** *The DAE and associated exosphere 550 hr post impact.* The sun is toward the bottom of the image. The DAE along the equator (*upper left*) is a narrow high-density region a few pixels wide (~6-8 pixels at 10.5 km/pixel) wide with a peak central density corresponding to the longitude of peak desorption. These desorbed molecules may hop toward the night side and adsorb onto the surface, or hop toward the day side (*upper right*) where they continue to migrate until they reach the terminator or are photodestroyed. *(Lower left)* this flow is pressure driven; the pressure peaks at the DAE, and decreases with distance. At this time, the DAE is optically thick (*lower right*), protecting most molecules within the DAE from photodestruction. We show the "photo factor", which is the local photodestruction timescale divided by the unshielded photodestruction timescale; this produces a quantity that describes how many times longer the local photodestruction timescale is compared to the unshielded photodestruction rate of 3.6 hours. The near-surface of the illuminated hemisphere is also well shielded from photodestruction, enabling significant migration from the dawn to dusk terminator. Note that the code treats the planet as "transparent" for the purposes of computing photodestruction timescales; the photodestruction timescales in Mercury's shadow are therefore an artifact of the numerical scheme implemented.

*3.3.2 Photodestruction*

The rate of photodestruction depends sensitively on the optical thickness (e.g., degree of self-shielding) within the plume and subsequent atmosphere. To compute the optical thickness of the atmosphere, we modify the procedure of *Prem et al. (2015)*, in which it is noted that 40% of $H_2O$ photodestruction is caused by Lyman-α photons, and 60% is caused by ultraviolet radiation with wavelengths between 145 and 186 nm (*Crovisier 1989*). As a result, *Prem et al. (2015)* found that the effective photodestruction rate ($r_{photo}$) is given by the expression

$$r_{photo} = r_{thin}\left[0.4e^{-nL\sigma_{121.6nm}} + 0.6e^{-nL\sigma_{145-186nm}}\right] \quad (7)$$

where $r_{thin}$ is the photodestruction rate when optically thin, $nL$ is the column density of water vapor (molecules/m$^2$), $\sigma_{121.6nm}$ is the Lyman-α absorption cross-sectional area (m$^2$), and $\sigma_{145-186nm}$ is the band-averaged absorption cross-sectional area (m$^2$). *Prem et al. (2015)* used data from *Watanabe and Zelikoff (1953)* to compute $\sigma_{121.6nm}$ to be $1.571\times10^{-21}$ m$^2$ and $\sigma_{145-186nm}$ to be $2.384\times10^{-22}$ m$^2$, assuming the Sun approximates a blackbody at 5780 K. We use this method to compute the reaction rate from column density, and then take the inverse of the photodestruction rate ($r_{photo}$) to get the photodestruction timescale. To understand how much this self-shielding increases the photo-destruction lifetime, we can divide this photodestruction timescale by the unshielded photodestruction timescale at 0.38 AU of $1.2\times10^4$ s, to compute a shielding factor that describes how many times longer the local photodestruction timescale is lengthened due to self-shielding. We present these computations in figure 10.

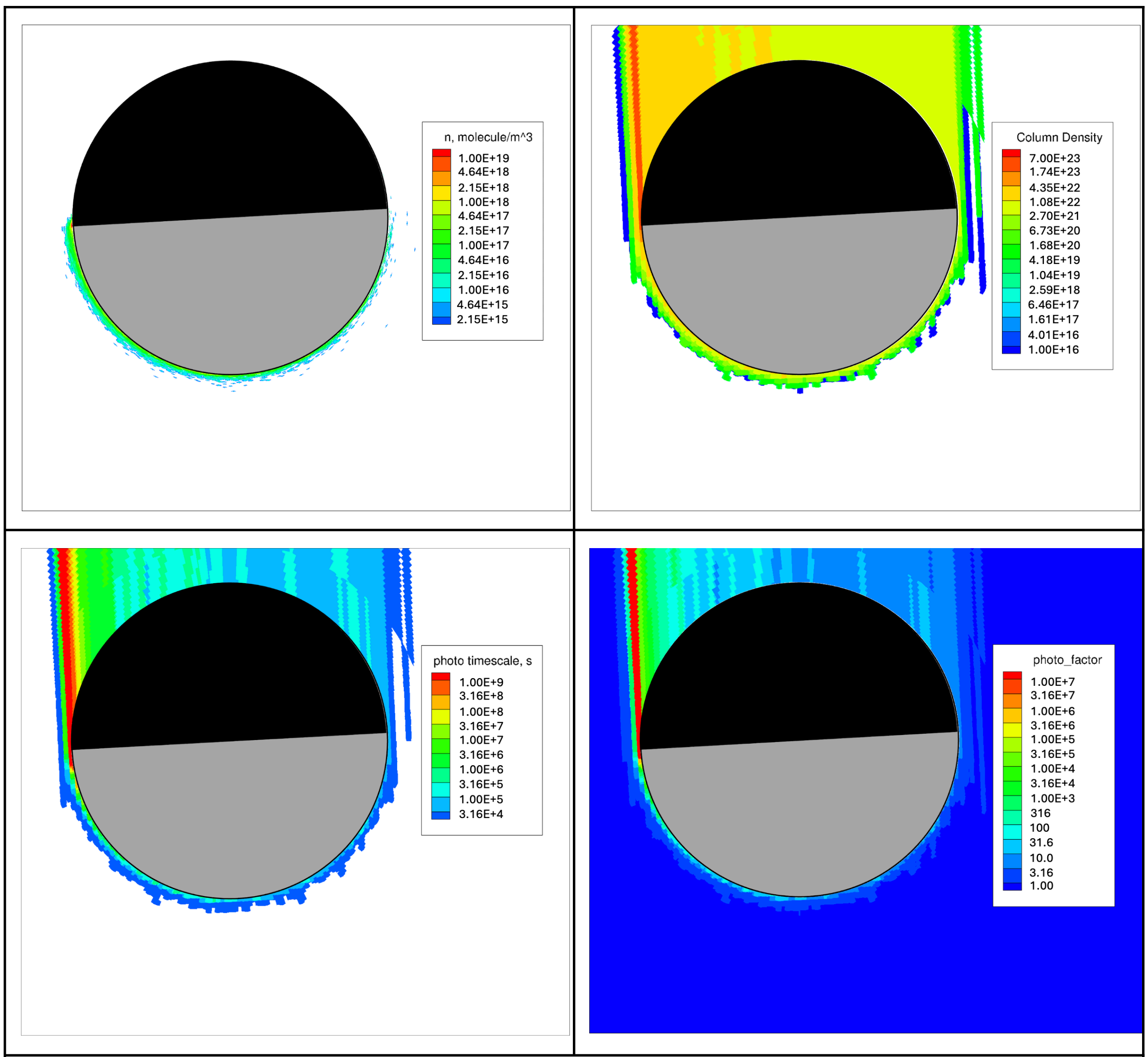

**Figure 10:** *Column density and photodestruction rate due to self-shielding at 500 hr post-impact.* At 500 hours post-impact, the simulation is in the quasi-steady phase (QSP), with the planet having rotated one quarter turn (Sun now toward bottom). (*Upper left)* The QSP exosphere extends across the illuminated hemisphere, shown here as a cross section along the equator with the atmosphere occupying the lower half of a circle above the surface of Mercury. The Dawn Atmospheric Enhancement (DAE), is seen here as a small cluster of red pixels on the left (dawn terminator). This relatively uniform distribution (varying only by a small factor with increasing distance from the DAE) leads to a similarly uniform photodestruction environment across the illuminated hemisphere; as column densities of $H_2O$ (*upper right*), and thus absorption of EUV photons increase toward the surface, the photodestruction timescale (*lower left)* increases dramatically by an order of magnitude, allowing water molecules to move along the lower illuminated hemisphere of Mercury toward cold traps or the terminator. (*Lower right*) These photodestruction timescales can be divided by the unshielded photodestruction timescale to determine the “photo factor” or multiplier by which the photo

destruction timescale has been extended.

We find that the DAE is sufficiently optically thick to shield most molecules behind it from photodestruction; the photodestruction timescale there is several decades. Thus, photodestruction at the DAE is negligible due to its limited spatial footprint. Molecules that are launched toward the night side hemisphere effectively travel ballistically from the DAE to their landing location where they stick to the surface. In contrast, if these molecules hop sunward of the DAE, the photodestruction rate increases significantly. Nevertheless, these molecules, which are now thermally migrating about the illuminated hemisphere are still largely shielded from photodestruction, as self-shielding from molecules higher in the exosphere increases the photodestruction timescale for the majority of these molecules by ~1.5 orders of magnitude. The resulting photodestruction timescale for these molecules increases to ~100 hours.

Interestingly, this timescale is comparable to the time required for these molecules to migrate to the opposite terminator. *Steckloff et al. (2022)* calculated that the average hop distance for a water molecule on Mercury is 60 km, and each takes on average 183 s to complete a single ballistic hop ($\tau_{hop}$). Using the random walk method described in *Steckloff et al. (2022)*, we can relate the average distance a particle has migrated from its initial starting point ($d_{migration}$) of a thermally hopping molecule to the average ballistic hop distance ($d_{hop}$) and total number of hops taken ($N$) since leaving the DAE:

$$d_{migration} = d_{hop}\sqrt{N}. \quad (8)$$

Rewriting this equation, we can derive the expected number of hops required to migrate a specified distance

$$N = \left(\frac{d_{migration}}{d_{hop}}\right)^2 \quad (9)$$

and multiply by the ballistic hop timescale ($\tau_{hop}$) to derive the migration timescale ($\tau_{migration}$), which is the expected timescale required to migrate by the specified distance ($d_{migration}$)

$$\tau_{migration} = N\tau_{hop} \quad (10)$$

$$= \tau_{hop}\left(\frac{d_{migration}}{d_{hop}}\right)^2 . \quad (11)$$

To compute the timescale required to hop to the far terminator from the DAE, we specify a characteristic length scale of that process (i.e., the radius of Mercury) for ($d_{migration}$); from this, we find that the migration timescale for a molecule to hop from the DAE to the far terminator is also on the order of ~100 hr. Thus, since the effective photodestruction timescale is comparable to the migration timescale, we would expect a significant number of molecules to both become photodestroyed and migrate to the cold traps or terminator. Indeed, this is exactly what is observed during the quasi-steady state, with nearly the same fraction of the initial molecules becoming photodestroyed (see section 3.3.1) as end up migrating into a cold trap (see section 3.3.3). This likely explains why the quasi-steady phase lasts for little over a single rotation, as the molecules present rapidly diminish to an optically thin density over this short time period.

The DAE during the quasi-steady phase is optically thick, effectively protecting the molecules within from photodestruction. This region of photo-shielding extends along the surface of the illuminated hemisphere, albeit at an ever-decreasing thickness with distance from the DAE. Nevertheless, sufficient shielding existed to enable significant numbers of molecules to reach the opposite terminator and cold traps. Photodestruction from insolation still occurs at the sunward side of the DAE and daytime atmosphere, albeit at a decreasing rate as photons penetrate deeper into these structures. By the end of the quasi-steady period, we find that the fraction of the initial molecules that were photodestroyed increases from 54% to 71.5% by ~110-120 days (~2600-2900 hrs) post impact.

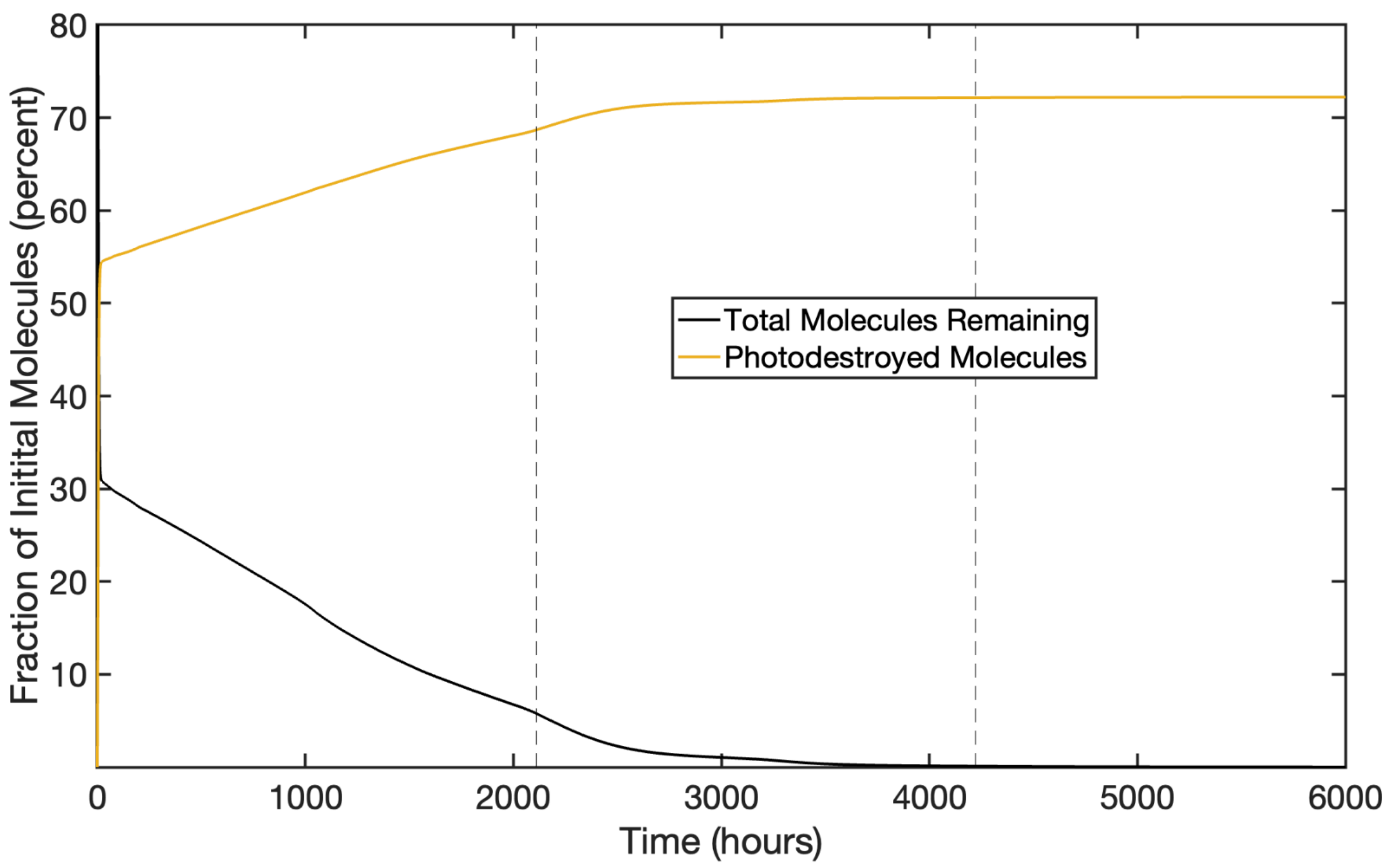

**Figure 11:** *Cumulative photodestruction during the quasi-steady phase.* At the start of the quasi-steady period (QSP), only ~31% of the initial molecules have yet to meet their fate (photodestroyed, escape, or migration to a cold trap); by the end of the QSP, only a few percent of the molecules remain. Additionally, 54% of all of the initial molecules were photodestroyed by the start of the QSP; during the quasi-steady period, an additional 17.5% of the initial molecules become photodestroyed. Full rotations of Mercury since the time of impact are denoted by the vertical dashed lines.

*3.3.3 Migration to the Cold Traps*

At the start of the quasi-steady phase (end of the reentry phase), only ~1% of the initial water vapor molecules resided in the cold traps. During the quasi-steady phase, this total generally increased linearly for ~110 Earth days (~2600 hrs), before leveling off at 14%. A major departure from this linear trend occurs at around 45 Earth days (~1100 hrs) post impact (one half of a complete rotation period), when the dusk terminator *at the time of impact* finally rotates into dawn. This occurs due to the enhanced frost density at the first sunrise at the terminator, which itself resulted from reentering material migrating to the terminators (see section 3.2.3). Thus, as this higher density of adsorbed water desorbs, a pulse of material is available for migration to the cold traps, causing the rate of cold trapping to transiently increase (see figure 12). An "echo" of this enhancement occurs again at ~90 days (~2200 hrs) post impact, when much of the water vapor from this initial enhanced desorption itself migrates to the dusk terminator, creating another longitude of enhanced water adsorption and corresponding increase in the rate of cold trapping (corresponding to the "bump" in figure 8, and a steepening of the curves in figure 12); *Prem et al. (2025, submitted*) observed a similar feature in comparable Mercury DSMC simulations. After this echo, there is little water vapor remaining adsorbed on the surface, and the amount of water that is cold trapped asymptotes to 14% of the original comet's water content, with approximately equal amounts of cold-trapped water in the northern and southern cold traps.

Although one would *a priori* expect more molecules to become trapped in the southern cold traps, which have a greater cumulative area, this southern area advantage is counteracted by the immediate distribution of ice. Significantly more ice was initially deposited proximal to the impact site (i.e., the north polar area) than more distal (figure 7). Furthermore, most water ice does not migrate far from its initial adsorption location on Mercury, moving on average approximately a dozen degrees across the surface unshielded (*Steckloff et al. 2022*), with shielding across the bulk of the illuminated hemisphere increasing this by a small factor. Thus, the water that ended up in the northern cold traps came predominantly from nearby initial ice deposits, which contained more ice in the north polar area than in the south. Thus, the cold traps in the north and south end up with approximately equal amounts of water.

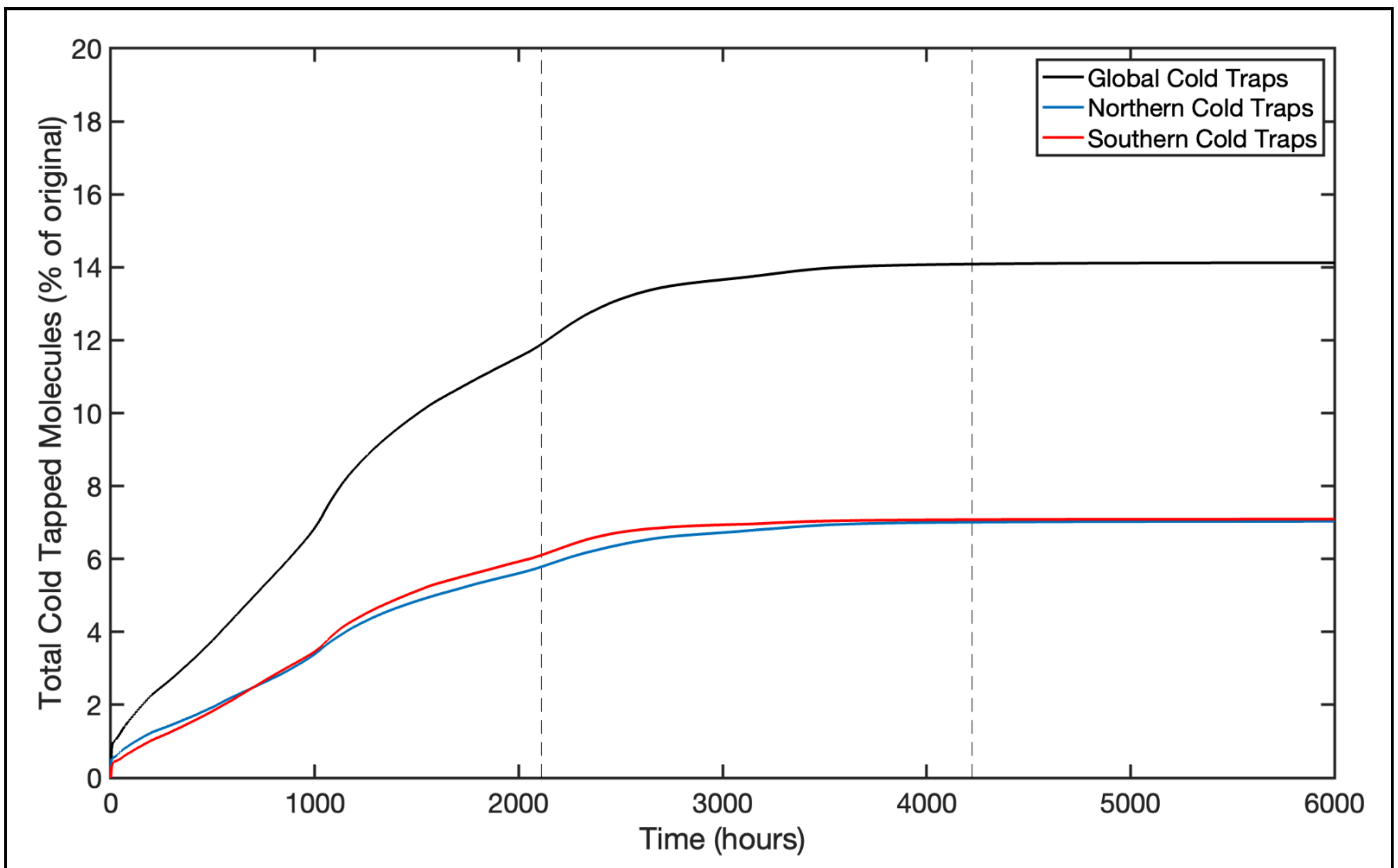

**Figure 12:** *Cumulative fraction of total molecules that end up in a cold trap.* The amount of cold trapped water vapor initially jumps to ~1% during the early and reentry phases. During the quasi-steady phase, the amount of cold trapped water vapor smoothly increases for the first ~110 Earth days post impact to asymptote at 14% of the initial amount of water. Black dashed lines denote complete synodic rotations of Mercury in our simulation. It appears that the asymptotic approach of the Southern and Northern lines to the same value is a coincidence only partly reflecting the relative cold trapping areas of the two regions.

We expected that there would be a longitudinal dependence of the mass flux into individual cold traps. Since the DAE is the location of greatest density, one would expect that a cold trap located along the DAE would experience enhanced deposition when its rotational phase aligns (i.e., temporarily lies along) with the DAE. However, we found no clear evidence of this longitudinal dependence on deposition. This may be due to the high latitudes of the cold traps, which are a significant distance away (much greater than the 60 km hop distance of these molecules; *Steckloff et al. 2022*) from the vast majority of molecules that enter the DAE at lower latitudes. Additionally, even an accessible nearby cold trap typically presents a very small target compared to the rest of the illuminated surface; this causes most desorbing molecules to enter the dayside exosphere to follow a random walk about the dayside. As a result, cold traps populate with exospheric water *whenever* they are on the illuminated side of the planet and interacting with the dayside high latitude exosphere, rather than just when located near the DAE.

### *3.3.4 The "Snowplow" Effect*

Desorbing molecules in the DAE follow a random ejection direction according to a Lambertian distribution. As a result, half of the molecules will hop back toward the unilluminated hemisphere, and readsorb onto the cold surface; this process was directly detected in Apollo 17 Lunar Mass Spectrometer data (*Hoffman and Hodges, 1975*). This increases the frost density immediately preceding the DAE. Over time as Mercury rotates, this causes an ever thickening deposit of frost to accumulate near the DAE, in a process analogous to a snowplow. This "snowplow effect" causes the amount of material in the DAE to increase as the initial frost on the night side hemisphere rotates into the DAE.

This produces significant effects on the volatile migration patterns. Since the DAE is already optically thick, feeding more material into the DAE has negligible impact on the photodestruction rate within the DAE itself. At the same time, this snowplow effect also causes a corresponding increase in flow toward the dayside as well, as molecules are as likely to hop to the dayside as they are to hop to the night side. This increases the density of the dayside exosphere, which leads to a corresponding increase in the rate at which material lands/deposits in Cold traps. Indeed, in figure 8, we can see that the cold trapping rate steadily increases until ~1100 hrs, when the initial dusk terminator at the time of impact finally rotates into dawn. This increased density also enhances the effective shielding from EUV photons and photodestruction. Thus, while the total rate of photodestruction (i.e., molecules per second) increases due to the Snowplow Effect, the *relative* rate (i.e., fraction of initial molecules photodestroyed per second) decreases, enhancing the efficiency of exosphere-mediated migration to Mercury's cold traps.

Finally, because the DAE and illuminated hemisphere are both in highly collisional/continuum regimes, this flow away from the DAE is pressure driven; that is, were a molecule on the dayside to try and "hop" toward the DAE, there are sufficient collisions (the mean free path is on the order of ~1 m in the DAE, and up to ~40 m near the top of the atmosphere) to stop nearly all molecules, and turn them around to flow toward regions of lower pressure.

#### *3.3.5 Effects of Rotation Rate*

The planetary rotation rate may be expected to play an important role in the survivability of water molecules on Mercury. Were the rotation rate to speed up, more molecules would be fed into the optically thick DAE per unit time, allowing more molecules to be protected from photodestruction then would occur at slower rotation rates. Conversely, at slower rotation rates, fewer molecules would be fed into the dayside exosphere via the DAE; thus the density of the dayside exosphere is inexorably linked to Mercury's rotation rate. Were Mercury to rotate faster, the enhanced density would increase the effective shielding within the exosphere, resulting in a reduced fraction in the number of molecules that are photodestroyed before reaching a cold trap (even though the absolute number of photodestroyed molecules will go up due to the sheer number of molecules in the exosphere) and enhancing the relative efficiency of cold trapping. Only in the limit that rotation is so slow to prevent the DAE from becoming optically thick would this effect be expected to disappear (this is itself the characteristic condition of the following Late-Phase; see section 3.4).

For a mean solar day on Mercury, this would result in twice as many molecules entering the DAE, doubling the density of molecules at the DAE, and thus increasing the amount of molecules migrating though the dayside exosphere. To zeroth order, these densities also roughly double. However, the increased density of water further enhances the effects of photoshielding, which inturn would enhance the number of molecules reaching either a cold trap or the oppositing terminator to “try again” on the next dawn. Combined, this would lead to a reduction in the amount of molecules that are photodestroyed, and an increase in the fraction of molecules that end up in the cold traps. Thus, these calculations should be taking as an upper limit in the photodestroyed fraction of water molecules, and as a lower limit in the fraction of cold trapped molecules.

No point on the surface of Mercury, however, experiences a mean solar day. Due to the 3:2 spin-orbit coupling of Mercury’s eccentric orbit, the Sun follows a complicated path across the sky that, at times appears to remain stationary and move retrograde near perihelion, leading to hot and cold poles across the surface. Thus, the conditions modeled in these simulations, are nevertheless relevant across the surface of Mercury at various parts of the orbit. However, simulating this complex motion is beyond the scope of this project.

*3.4 Late Phase*

As the adsorbed frost thins due to photodestruction and migration to the cold traps, few mobile molecules remain after ~1.5 rotations (~150 Earth days post impact), with ~0.2% of the initial molecules yet to reach their ultimate fates. At this point, the DAE and dayside exosphere become optically thin, which defines the start of the late phase and the end of the quasi-steady phase (QSP). This condition allows EUV photons to rapidly destroy any remaining molecules. As a result, only minor additions to the cold traps occur during the late phase, with less than ~0.1% of the initial molecules moving to cold traps throughout the late phase, and the majority becoming photodestroyed. The late phase is the final phase of plume evolution and hardly contributes any substantial mass to the cold traps.

This 1.5 rotations timescale is likely not fundamental to Mercury, but rather to the amount of frost initially deposited on the surface. Thicker frost layers would lead to denser DAEs and enhanced dayside shielding, leading to a lower loss rate. In turn, this would facilitate greater numbers of molecules reaching the dusk terminator to “try again tomorrow morning.” Such differences in frost thicknesses could result from different impacts; larger impacts should deposit more material due to both possessing more material and to enhanced shielding resulting in lower loss rates. Similarly, slower impacts should also lead to relatively thicker initial frost deposits as less material would be lost ballistically. However, at sufficiently slow speeds, the geographic distribution of this frost may differ significantly from what we modeled. This timescale of ~1.5 rotations is also consistent with what was observed in previous studies of similar comet impacts into the Moon (*Stewart et al. 2011*), suggesting that this may be characteristic timescale for clearing transient frosts resulting from 1 km radius comet impacts from the surfaces of airless bodies in our solar system. Nevertheless, exploring this size dependence is beyond the scope of this work.

*3.5 Fates of Molecules*

For the presently assumed conditions (1 km radius comet nucleus, 30 km/s, 60 degree impact angle), we calculated the ultimate fates of the water molecules released by the comet impact (figure 13). We find that 65% of the initial comet water mass (ICWM) molecules had sufficient energy to reach Mercury's Hill sphere and escape the system, however the vast majority of these (79%, or about 51% of the ICWM) photodestruct prior to reaching the Hill sphere; 13% of the ICWM escape to the Hill sphere intact. In spite of high surface temperatures that can impart significant thermal velocity into the vapor molecules, Mercury's high escape speed of 4.3 km/s precludes additional ballistic loss. Thus, the remaining 37% of the molecules in our simulation (i.e., 37% of the ICWM) that remain bound to Mercury either photodestruct or migrate to the cold traps.

Of the initial molecules, 23% of the ICWM are photodissociated while bound to Mercury, with 20% of this photodestruction (~4.6% of the ICWM) occurring during the initial expansion and collapse of the plume in the first ~24 hrs post-impact. Once the plume settles into the quasi-steady phase (QSP), the rate of photodestruction slows significantly, with the remaining 80% (~18% of the ICWM) photodissociating over the ensuing ~100 days. This process is likely to release significant quantities of water-group ions ($O^+$, $OH^+$, $H_2O^+$) to Mercury's exosphere.

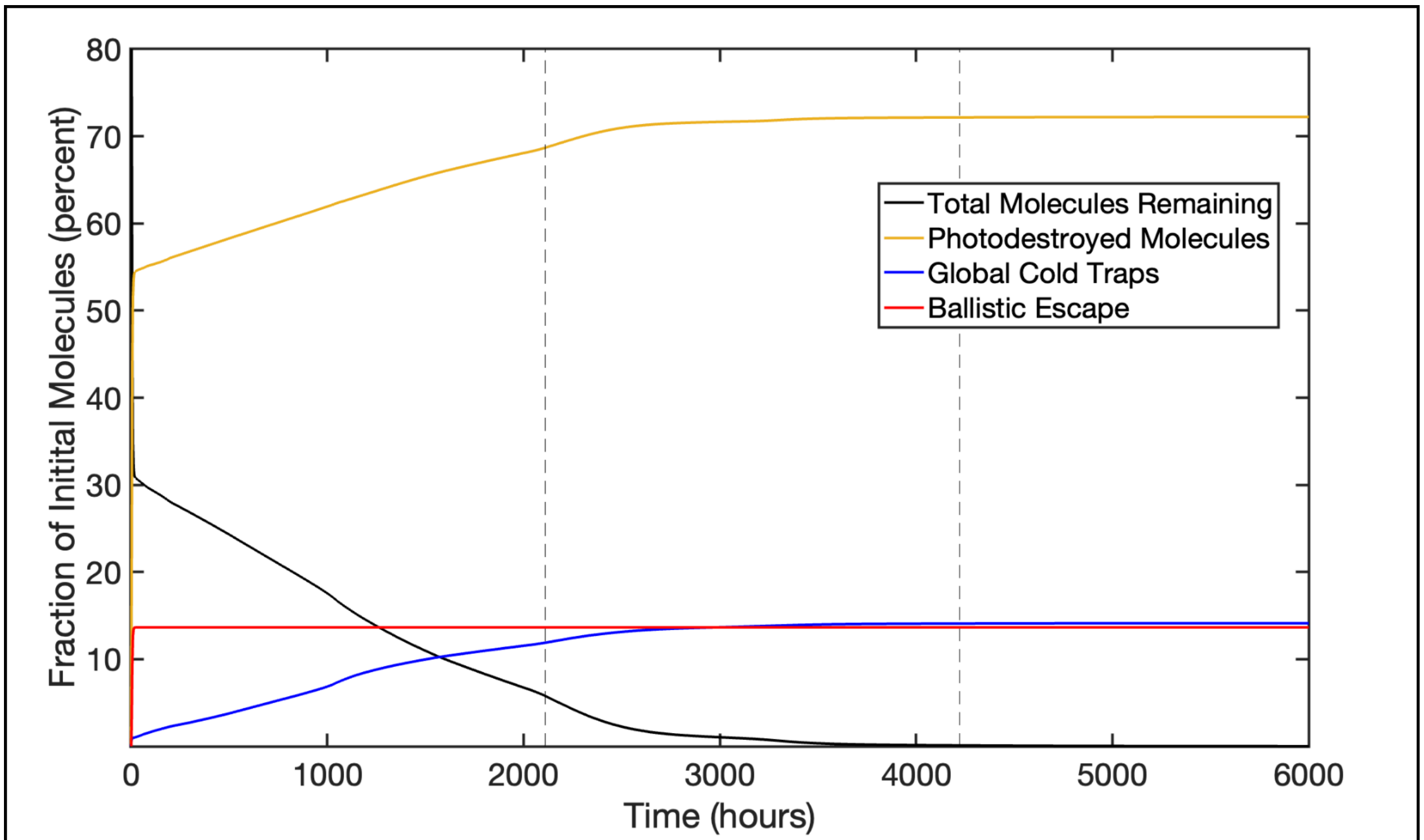

**Figure 13:** The eventual fates of the simulated molecules in our simulation over time. Although the majority of molecules escape the system, only ~13% reach the Hill sphere without photodestruction (which PLANET records as ballistic escape). 14% of the molecules migrate to the cold traps, with the remaining 73% photodestructing prior to reaching a cold trap or escaping.

This exosphere-mediated migration of water to Mercury's cold traps lays down a surprisingly thick layer of ice in the cold traps often equivalent to several mm of ice from a 1 km impactor. In the northern hemisphere, the thickest deposits of water into a cold trap in our simulation were the few cm of ice in the impact craters closest to the impact site; in general, cold trap deposit thicknesses decrease with increasing distance from the impact site. A similar trend was observed in the diffuse cold traps, with the highest average areal density found in the 80ºN-90ºN cold trap nearest to the impact site (190g/m$^2$), decreasing to 44.2 g/m$^2$ and 16.0 g/m$^2$ in the more distal 70ºN-80ºN and 67ºN-70ºN cold traps respectively. In the southern hemisphere, however, this pattern reverses, with the more distal (from the impact site) 80ºS-90ºS diffusive cold traps hosting 308 g/m$^2$ compared to the more proximal 73ºS-80ºS diffusive cold trap; this suggests that antipodal focusing and the latitude-based trapping probabilities dominate any effect on impact distance in the ultimate distribution of cold-trapped water. Ultimately, these diffusive cold traps were the dominant stores of cold-trapped water, with 33% and 39% of the total cold trapped water landing in, respectively, the Northern and Southern diffuse cold traps. For a plot of deposit areal *density*, see figure 14).

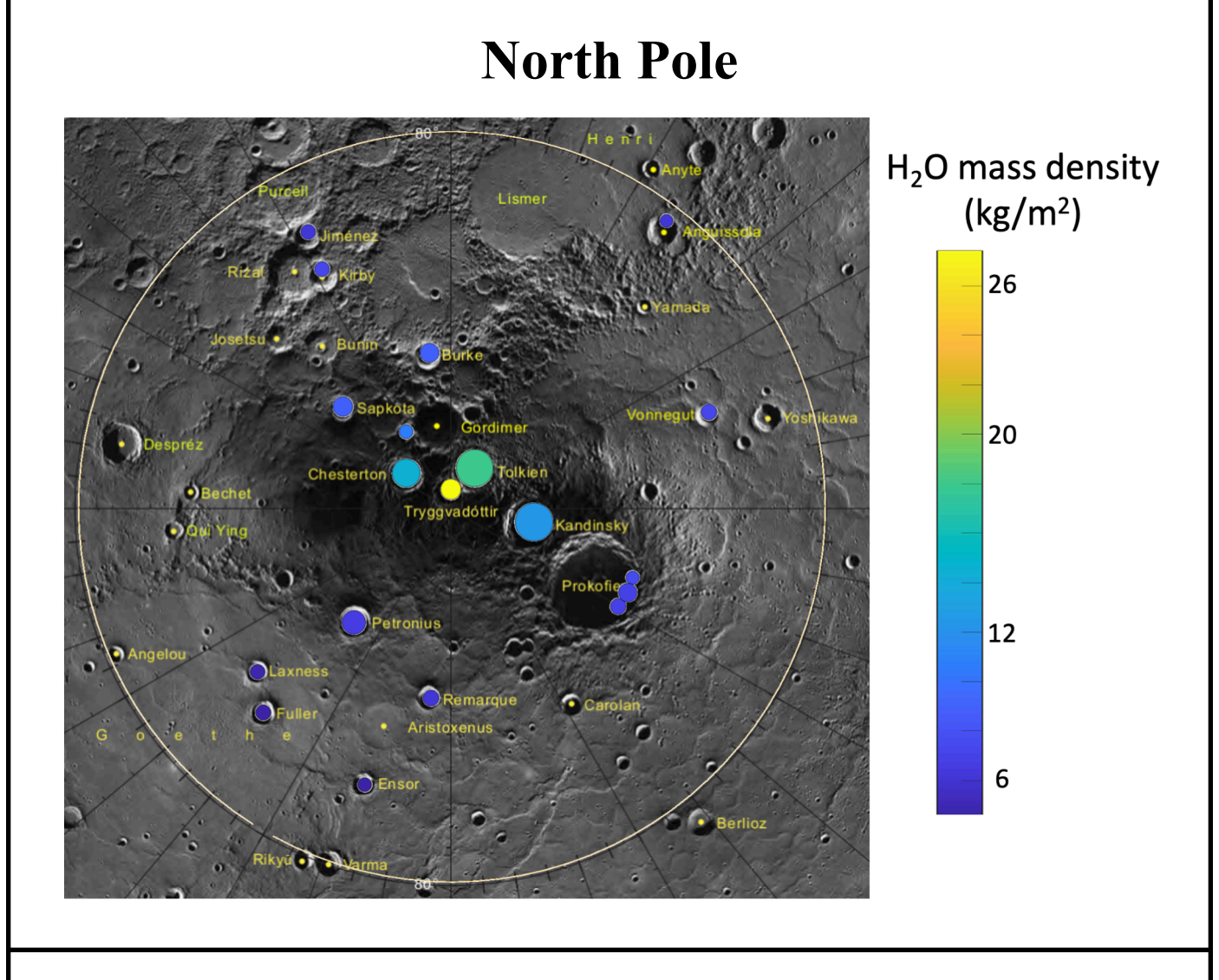

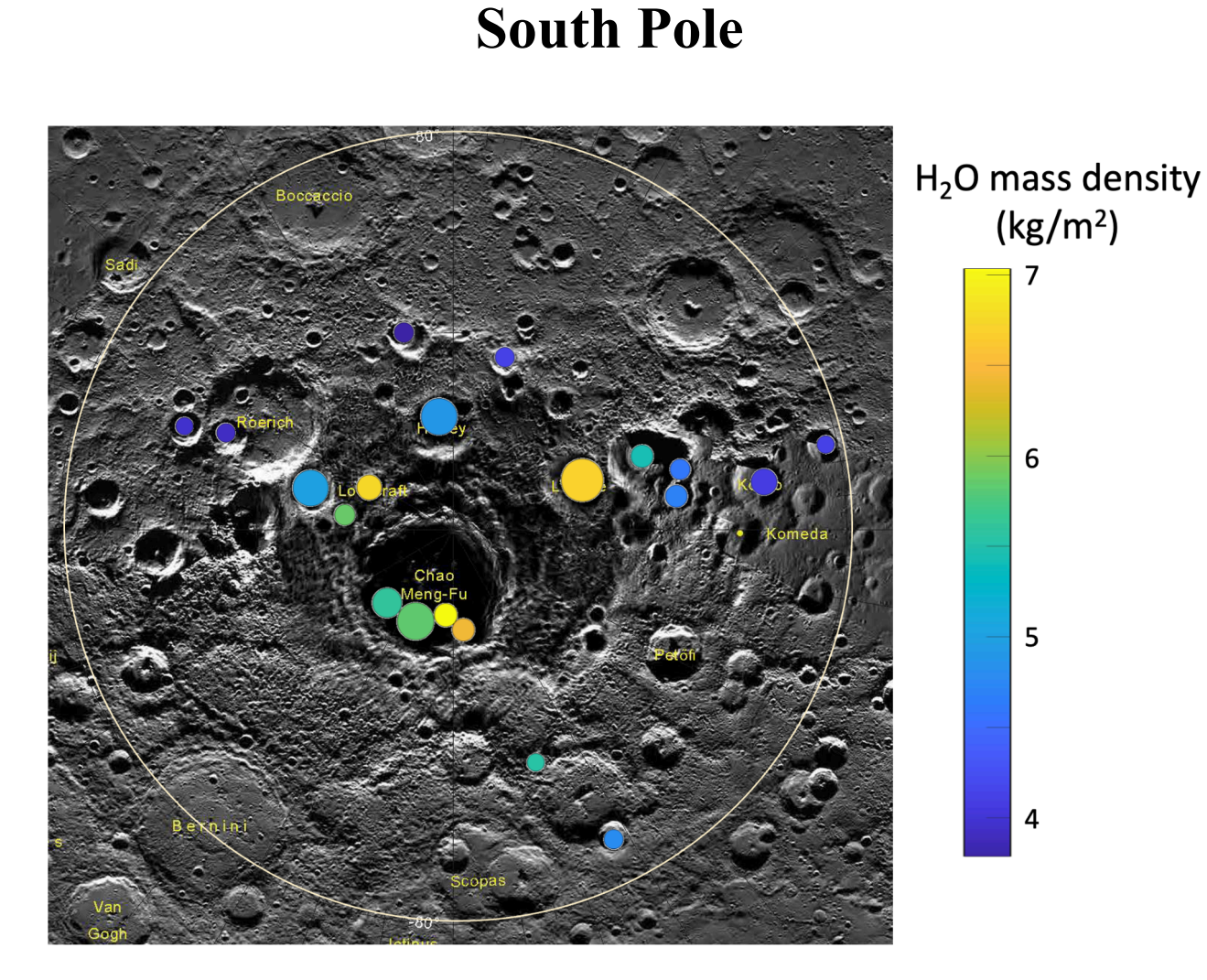

**Figure 14:** *Mass density of deposited ices into Mercury's discrete cold traps.* The comet impact deposits significant water into Mercury's cold traps; The discrete cold traps collect (*top*) ~6-26 kg/m$^2$ in the North and (*bottom*)~4-7 kg/m$^2$ of comet ice throughout the impact process; with cold trap densities generally increasing with increasing latitude; highest densities are found closest to the impact site at the north pole. The sizes of these circles correspond to their surficial extent. Note different scale bars for north and south poles. Diffuse cold traps are not plotted in these figures, as they would cover the discrete cold traps. Cold trap densities plotted above a subset of the USGS maps of the Borealis quadrangle in the north *(top)* and the Bach Quadrangle in the south *(bottom)*

## IV. Discussion

The fate of comet impact-liberated water depends on competing processes operating on timescales that are sensitive to, among other things, a body's orbital distance from the sun (photodestruction), gravity (ballistic escape), and comet impact speed (retained fraction). As a result, these processes are likely to behave quite differently on each object in the Solar System, and will likely depend on which processes dominate. For the case of Mercury itself, considerations of differing processes, geometries, and conditions have led to considerable variation in the amount of water that impacts are expected to be able to deliver. For example, *Moses et al.* (*1999*) found micrometeorites may have delivered (3-60)$\times 10^{13}$ kg of water ice to Mercury's cold traps over the past 3.5 Gyr (equivalent to an ice thickness 0.8-20 m thick) however, subsequent micrometeoritic erosion could reduce this value by half (Moses et al. 1999). They further broke these potential contributions by source population, with Jupiter Family Comets (JFCs) contributing (0.1-200)$\times 10^{13}$ kg of water to Mercury's cold traps (0.05-60 m of ice), and Halley Family Comets (HFCs) contributing an additional 0.2-20$\times 10^{13}$ kg of water (0.07 - 7m of ice). Nevertheless, they note that the cleanliness of the deposits suggest delivery from a few large impacts, which could also explain the quantity of water in the cold traps *(Moses et al 1999)*.

Our results suggest that any single comet that delivers a significant fraction of the $10^{13}$-$10^{14}$ kg of cold trapped water would have to be significantly larger than the 1 km-radius comet considered in this study (total water mass of ~$10^{12}$ kg) which we found could deliver only a few ~mm-cm of ice to the cold traps. However, a ~10 km comet nucleus would contain 1000 times more water, and thus could plausibly populate Mercury's cold traps assuming the percentages of the initial comet water mass (ICWM) are subject to photolysis, ballistic escape, cold trapping, and photodestruction are the same as our 1 km radius comet nucleus. In fact, the increased amount of material in a larger impactor could dramatically increase the degree of photolytic shielding, protecting significantly more water from photodissociation. Thus a smaller impactor (perhaps a half order of magnitude smaller than this ~10 km estimate) could provide the amount of water needed to populate the cold traps. Indeed, the 95 km diameter Hokusai Impact Crater (*Harmon et al. 2007*) appears to have been created by an impactor approximately 6.5 km across (using the Purdue Impact Calculator (https://www.eaps.purdue.edu/impactcrater/crater_c.html), assuming a 30 km/s impact with typical comet density of 530 kg/m$^3$ (*Pätzold et al. 2016*; Steckloff et al. 2021 and references therein), Hokusai is a relatively young crater, having formed ~100 Mya (*Neish et al. 20123; Ernst et al. 2018*), which aligns nicely with the <330 Ma age of the polar ice (*Deutsch et al. 2019*). This implies thatleading to the notion that the Hokusai impactor could have populated Mercury's cold traps (*Ernst et al. 2018),* an idea further supported in *Prem et al. (2025*). This would be comparable to the size of the nucleus of 9P/Tempel 1 (*Ahearn et al. 2005*), a typical JFC.

For their part, *Ernst et al. (2018)* estimated the amount of water delivered to Mercury's cold traps by scaling to Mercury the amount of water that *Stewart et al. (2011)* calculated could be delivered to the lunar cold traps. This is a particularly compelling comparison, as *Stewart et* *al. (2011)* used an earlier version of the PLANET DSMC code (lacking photolytic shielding) to investigate how much water would migrate to the cold traps from a 1 km-radius comet striking the Moon at 30 km/s, albeit at an impact angle of 45°. *Stewart et al. (2011)* found that ~3.1% of this comet's water content would be initially retained by the Moon; from this, *Ernst et al. (2018)* estimated that 45-50% of the initial water molecules would be retained by Mercury's gravity; this is consistent with the initial state of our simulations (30s post-impact), in which 45.5% of the water molecules remained gravitationally bound to Mercury (see section 3.1 on the early plume). However, this result neglects subsequent gas-driven acceleration of the plume, which reduces the retention rate to 35% (see section 3.1). *Ernst et al. (2018)* further assumed that 10% of the retained water would migrate to the cold traps, using the mean of the ~5-15% migration efficiency on Mercury determined by *Butler (1997),* the only such relevant study available at the time; this results in an overall delivery efficiency from comet to cold trap of ~5%. This is approximately a factor of 3 smaller than the 14% we derive here.

Because of limited data at the time, *Butler (1997)* assumed the same cold trapping area fraction as the Moon. To explore if this factor-of-three difference is due to photoshielding or differences in the cold trapped areas used in our respective calculations, we use updated results since *Butler (1997)* to improve this comparison. The models of *Butler (1997)* and *Schorghofer et al. (2017)* are quite similar, with the model of *Schorghofer et al. (2017)* in particular known to produce essentially the same results as PLANET DSMC, when run with the same inputs in Free Molecular mode (no collisions) without shielding. *Schorghofer et al. (2017*) found that, for a cold trapped area fraction of ~0.02%, approximately 1% of the initial molecules will migrate to a cold trap; if we scale this result by the cold trapped area fraction in our simulation (equivalent to ~0.075% the surface of Mercury), we obtain a trapping efficiency of ~4%. This result agrees well with the 5% lower limit calculated by *Butler (1997)*, but would skew the estimate of *Ernst et al. (2018)* even lower, to only a ~2% migration efficiency for impact delivered water vs. the 14% we compute with comparable conditions (except for photolytic shielding). This suggests that the effect of photolytic shielding posited by *Ernst et al. (2018*) and described by *Prem et al. (2015)* could enhance the survival of migrating water to the cold traps from comet to cold trap for a 1 km-radius comet by nearly an order of magnitude; for a Hokusai-sized impactor this shielding is likely even more effective. This suggests that photoshielding is a dominant effect, and must be included when considering the effects of volatile migration on Mercury.

The amount of material that is initially gravitationally retained during an impact into Mercury has also been explored. *Moses et al. (1999)* computed the fraction of impacted material that would be gravitationally retained, assuming a hot, isentropic free expansion of a vapor plume into a vacuum. For this, they use the procedure of *Vickery and Melosh (1990)* to calculate the specific internal energy of the expanding vapor plume mode for a 45° impact. Using this model, *Moses et al. (1999)* calculate that only ~9% of the initial vapor would be retained. Although this model accounts for vapor (pressure-driven) expansion, we find that it grossly underestimates the amount of material initially retained by a factor of 4. This may be partly due to the differences in impact angles, as *Moses et al. (1999)* include vaporized target material in the calculation, and note that the *total* mass of vaporized vaporized material ($M_{vaporized}$) scales as

$$M_{vaporized} \sim v^2 cos^4\theta \quad (12)$$

where $v$ is the impact speed and θ is the impact angle (*Moses et al. 1999*). Merely a difference in impact angles may explain some of this difference, as the 60° impact we consider would be expected to produce exactly 4 times as much impact vapor, according to this formula, producing the quantity of material that we calculate. However this formula includes vaporized target material, whereas we consider only vaporized impactor material (we assume the target to be dry); thus it remains unclear how much of this discrepancy between our respective results is due to differences in impact angle vs. other fundamental assumptions.

Relatedly, *Bruck Syal et al. (2015)* explored the amount of material retained on Mercury as a function of both impact speed and target porosity. Our computation assumes a comet nucleus to be a solid sphere of water ice (zero porosity, density of 920 kg/m$^3$), when in reality, comet nuclei have high bulk porosities of ~75% (*Brouet et al. 2016*; *Henrique et al. 2019*) aggregates of ice and dust, with densities typically ~500 kg/m$^3$ and contain significant quantities of refractory materials (*Pätzold et al. 2016; Steckloff et al. 2021*). *Bruck Syal et al. (2015)* considers a smaller, 1 km-diameter nucleus that was treated as a *porous* water ice sphere, with considered porosities varying from 0% to 60%. Because of how impedance matching conditions depend on these impact parameters, objects that are closer in density tend to have better transmission of impact energy. Conversely, widely differing densities between impactor and target would exhibit poorer energy transfer, and less material reaching escape speed (*Bruck Syal et al. 2015*); this is consistent with the $SiO_2$ target and porous water ice considered by *Bruck Syal et al. (2015)* colliding at 30 km/s, which they find results in only 16% water being retained. Exactly how much water would be retained were we to consider porous comets is presently unclear; nevertheless, it is likely that our results represent an upper bound of how much material would be gravitationally retained, given the high bulk comet density that we consider.

Finally, *Ernst et al. (2018)* also scaled the work of *Ong et al. (2010),* to estimate the retention of water on Mercury. *Ong et al. (2010)* studied volatile retention from comet impacts into the Moon using the SAGE hydrocode and the Pactech/SAIC tabular Equation of State for water ice (which contains six phases of water ice, in addition to vapor and gas; *Ong. et al. 2005)*. Comets were modeled as solid spheres of water ice; at 30 km/s, *Ong et al. (2010)* found that a little over 1% of the initial comet mass is retained. *Ernst et al. (2018)* scaled this result for Mercury, to find that ~24% of the impacted water is retained by Mercury. This is approximately ⅔ of what we found, but the agreement is good given the differences in underlying hydrocode structure, equations of state, etc.

*4.1 Comparison with Similar Impacts into other Airless Bodies*

The PLANET DSMC Code, in this configuration, has been used to look at an identical comet impact into *all* of the other large airless bodies in the inner Solar System: Mercury, The Moon, Ceres, and Vesta (*Prem et al. 2015; Steckloff et al. 2018*). In this section, we compare these results for Mercury with those for the Moon, particularly the results of *Prem et al. (2015)*, which are most relevant to this work. We also compare these results with Ceres and Vesta.

*The Moon*

*Prem et al. (2015*) modeled a similar impact (1 km radius, 30 km/s, 60°) at the Moon using the same PLANET DSMC code, allowing for a direct comparison of the behavior between identical comet impacts into the Moon and Mercury. This is a crucial comparison for understanding how the relatively gentle photodissociation environment and reduced gravity of the Moon affect its impact-liberated exosphere. Although some trends are expected, (e.g., Mercury's higher gravity favors greater water retention and more compressed/narrow atmospheric structures), these differences can interact with one another in non-intuitive ways.

In the initial moments, these impacts are in the hydrodynamic regime, where the evolution of the material is determined solely by the impact energy. Over time, the differences between Mercury and the Moon become apparent. In the earliest phases of the impact plume's evolution (30s post impact), only ~28% of the initial water molecules were gravitationally bound to the Moon, versus 46% retained by Mercury at the same point. Such a stark difference is not surprising given that Mercury's surface escape speed of 4.3 km/s is close to double the lunar surface escape speed of 2.4 km/s. The speed distribution of the initial molecules is very broad (see figure 1 for distribution of water molecule speeds at 30s post-impact.) However, although more water is initially gravitationally bound, the rate of photodestruction is also significantly higher at Mercury.

At first glance, one might naively assume that a comparable fraction of additional molecules would be retained by Mercury to participate in the subsequent evolutionary phases (reentry shock atmosphere and migration to the cold traps). However, any additionally retained molecules are still exposed to the extreme EUV photon environment around Mercury. This increase in retention would occur at the higher-speed margin of the impact speed distribution, where the additionally retained molecules would be on trajectories that leave them exposed to solar radiation for a considerable amount of time. Thus, one should expect these additional molecules to be photodestroyed considerably faster than they would at the Earth. Indeed this photodestruction of the impact vapor plume at Mercury is so severe that the density of the reentry shock atmosphere is nearly the same for both the Moon (*Prem et al. 2019a*) and Mercury (see figure 5): the density at the surface is ~$10^{18}$ molecules/m$^3$, dropping to ~$10^{17}$ molecules/m$^3$ just below the reentry shock layer. This is expected behavior for an atmosphere, which should exponentially decrease in density with increasing altitude.

However, the shock atmosphere of Mercury shows the effects of Mercury's higher gravity relative to the Moon, with the force of the infalling gas and hydrostatic forces of the upper atmosphere compressing the day-side atmosphere. A stable atmosphere should exponentially decrease in density over a length scale determined by the atmospheric scale height (*H*), which measures the height over which the atmospheric density should drop by a factor of 1/*e* (see equation 6). For the Moon, the scale height of its atmosphere at 300 K is 85 km, which is approximately half of the observed thickness of the day side atmosphere. Indeed, atmospheric density drops by approximately an order of magnitude over this height (*Prem et al. 2019a*), which is consistent with expectation. For Mercury, the scale height is only ~40 km, roughly half the height of its shock atmosphere. Over two scale heights, the density should drop by $1/e^2$ (0.135). Indeed, Mercury's shock atmosphere's density also decreases by a full order of magnitude from the bottom to the top (figure 5). On Mercury, this reentry shock atmosphere is ~86 km, whereas it is ~200 km on the Moon; these atmospheres have essentially the same atmospheric structure, with Mercury's chief difference being an atmospheric thickness only ~⅖ the height of the equivalent lunar atmosphere.

Given that both Mercury and the Moon exhibit atmospheres that have the same density boundaries and structure, the ratio of the instantaneous masses of these respective atmospheres is merely the ratio of their respective volumes (i.e., surface area times height). From this, we find that this ratio is ~1, suggesting that, to first order, these two atmospheres initially have the same mass. This is further evidence that enhanced photodestruction at Mercury offsets the additional gravitational retention of water molecules. However, this result is likely sensitive to Mercury's location in its orbit, whose eccentricity of 0.205 and semimajor axis of 0.387 AU result in a perihelion distance of 0.307 AU and aphelion distance of 0.467 AU. This corresponds to an EUV flux that is 145.6% and 62.9% the EUV flux assumed in this paper at perihelion and aphelion, respectively. The resulting differences in the $H_2O$ photodestruction timescales between perihelion and aphelion may lead to equally significant differences in the amount of material that is photodestroyed prior to reentering the atmosphere.

Regardless, Prem et al. (*2015*) found that only ~½ percent) of the initial cometary water migrated to the lunar cold traps, versus 14% for Mercury (see section 3.5). However, there were some important differences in simulation set-up:

1. The simulations in Prem et al. (*2015*) were run for only 5 Earth days (since the focus of the work was on exploring the structure of the impact-generated atmosphere), and as a result the migration of water to cold traps was not modeled to completion.
2. Prem et al. (*2015*) modeled only a subset of polar PSRs (as in Stewart et al., *2011*), which covered a total area of 5831 km$^2$, considerably less than the total lunar cold trap area of 28,921 km$^2$ poleward of 80° (Mazarico et al., 2011).
3. These earlier Prem et al. (*2015*) simulations did not use the radiative heat transfer methods introduced in Prem et al. (*2019*), which may influence the distribution of water between cold traps.

We address (1) as follows: At the end of 5 Earth days, a total mass of $1.91\times10^{10}$ kg was cold-trapped, and $6.17\times10^{11}$ kg remained in the exosphere or adsorbed to the surface outside cold traps. Based on longer-term collisionless simulations using the same code (*Prem et al., 2018*), we can conservatively estimate that ~5% of the mass remaining would ultimately migrate to cold traps after several lunar days, adding an additional (5% of $6.17\times10^{11}$ kg) = $3.09\times10^{10}$ kg and bringing the total cold-trapped mass to ($1.91\times10^{10}$ + $3.09\times10^{10}$) = $4.99\times10^{10}$ kg, or an average areal density of 8.57 kg/m$^2$ over the modeled cold trap area - comparable to the densities found here for Mercury.

If we scale the cold-trapped mass to account for the larger than modeled cold trap area by Prem et al. (*2019*), we obtain an updated estimate of ($4.99 \times 10^{10}$ x 28,921/5831) = $2.48 \times 10^{11}$ kg cold-trapped, equivalent to ~5% of the comet mass. This is less than the ~14% estimate for Mercury, likely due to a combination of higher gravity and larger cold-trapping area, with self-shielding mitigating against higher photolytic loss rates at Mercury.

*Ceres and Vesta*

We attempted to repeat this study for both Ceres and Vesta as previous studies have found that loss is dominated by ballistic escape, rather than photodissociation as is the case at Mercury and the Moon. We scaled the initial 30s post-impact state of the plume to Ceres and Vesta, and ran the DSMC simulation. However, our simulations quickly revealed that nearly all vapor molecules are lost ballistically. Indeed, an inspection into the initial state of our simulation reveals that Ceres and Vesta would retain only 0.032% and 0.0077% of this initial water vapor, respectively; this is equivalent to a layer spread across the surface only ~120-150 molecules thick. Since Vesta has no known cold traps (*Stubbs and Wang, 2012*), these molecules would be rapidly lost. Based on their loss rate per ballistic "hop" of 79% (*Steckloff et al. 2022)*, these molecules would be lost to space after ~3-4 hops.

On Ceres, the greater gravity drives greater retention, with a loss rate per ballistic "hop" of 48%, which is equivalent to each molecule experiencing an average of 1.1 hops before being lost to space (*Steckloff et al. 2022)*. Multiplying this expected number of hops by the thickness of the layer, we expect each molecule on the surface should experience a total of ~165 molecular hops, before the entire volatile layer is lost to space. Given the near-global extent of Ceres exosphere, these molecules can be thought of as coming back randomly across the surface to a good approximation. At its current axial tilt, the Cerean permanently shadowed regions (PSRs) occupy ~0.13% of Ceres' surface area (*Schorghofer et al. 2016*), a small fraction of Ceres' total area. Thus, these cold traps would trap a layer ~165 molecules thick before the exosphere dissipates; this would be in addition to the ~150 molecule-thick layer that would be directly deposited following the impact. This suggests that 30 km/s impactors themselves are a poor source of water for the Bright Crater Floor Deposits (BCFDs) observed on Ceres, a perhaps obvious result. However, this does neglect any water contribution to the cold traps from Ceres' water-rich mantle, which could produce significant amounts of additional water-rich ejecta that could amplify the amount of water delivered to the cold traps. Nevertheless, the effects of such ejecta are beyond the scope of this study.

*Scaling for Different Impact Speeds and Impactor Sizes*

Crater and ejecta scaling laws enable the behavior of a single impact event to predict the effects of different impact conditions. These scaling laws take advantage of the approximate point-source origin of the impact energy (akin to an underground nuclear explosion) as distance from the impact point increases (*Melosh, 1989; Housen and Holsapple, 2011*). This results in scale-independent solutions to the impact cratering process (*Housen and Holsapple, 2011*) via a series of non-dimensional scaling constants (*Melosh, 1989*). In the case of the generation of vapor from a comet impact into Mercury, the ejecta of interest that populates these plumes originates from the impactor itself, and is therefore ejected very near the impact site. Nevertheless, ejecta scaling laws have been found to be accurate as close in as one impactor radius from the impact point (*Housen and Holsapple, 2011 and references therein*). This suggests that these crater scaling laws may be sufficiently accurate for the purposes of this discussion.

Our impact into Mercury lies within the gravity-dominated regime, where the ejecta velocity ($v$) scales as

$$\frac{v}{U} \sim \left(\frac{x}{a}\right)^{\frac{-1}{\mu}} \tag{13}$$

where $U$ is the impactor speed, $x$ is the distance from the impact point to the launch location (as measured at the plane of the original ground surface), $a$ is the impactor radius, and μ is material property that depends on the high-pressure behavior of the material (*Housen and Holsapple, 2011*). Although it is unclear where this ejecta pierces the plane of the original ground surface, it is unlikely to vary significantly; in such a case the right-hand-side of this expression becomes a constant, and the velocity of the ejecta ($v$) becomes linearly proportional to the impactor velocity ($U$). In principle, this means simply scaling the vapor speeds in figure 1 to different impactor speeds.

Immediately, we see that fewer and fewer vapor molecules are gravitationally retained at greater impact speed; at speeds of ~60 km/s, the peak of this distribution would be at the escape speed of the Mercury system, with vanishingly few vapor molecules retained at speeds above this. Energy conservation shows that a typical Oort Cloud comet would have a speed of ~68 km/s at the distance of Mercury's orbit, suggesting that typical Oort Cloud comets would result in little retained mass and water delivered to the cold traps. However, Mercury orbits the Sun at ~48 km/s; although head-on collisions (impact speed of 116 km/s) would retain negligible impactor mass, impact speeds would only be ~20 km/s if Mercury were "rear-ended", and the results here would be applicable to this latter subset of Mercury impacts. Neglecting the effects of gravitational focusing, we can use trigonometry to estimate the angles between the comet and Mercury's respective velocity vectors (This term chosen to avoid using "impact angle", which traditionally describes the angle the impactor makes with respect to the surface it strikes) required to generate impacts within 10 km/s of the 30 km/s impact we simulated. We find that a 30 km/s and 40 km/s impact would have a velocity vector offset from Mercury's by 23$^{o}$ and 35$^{o}$ respectively. Given the isotropic distribution of Oort Cloud comets, approximately 9% of the Oort Cloud comets that strike Mercury would hit with an impact speed of 30±10 km/s, for which our results would be most applicable.

Using the vis-viva equation, we can also compute the orbital speeds of Jupiter Family Comets (JFCs), which have aphelia near Jupiter and perihelia (for those that strike Mercury) near Mercury, which are only slightly slower at ~62 km/s. Unlike the Oort Cloud Comets, JFCs tend to occupy a small range of orbital inclinations of ~±35° from the ecliptic (*Di Sisto et al. 2009*); this largely precludes head-on impacts in favor of JFCs rear-ending Mercury. Thus, actual impact speeds would tend to be significantly lower for JFCs, perhaps ~14-35 km/s. Thus the results of our simulation are most analogous to JFCs striking Mercury, and suggest that, all else equal, such JFC impacts may dominate the delivered flux to Mercury's cold traps.

Although we can scale these results for different impact speeds and amounts of retained material, our work shows that this retained material tends to interact with itself (e.g., self-shielding) to enhance the amount of water retained and ultimately migrated to the cold traps. Computing the effects of this self-shielding is not *a priori* simple to scale, as the rate of photodestruction depends non-linearly on retained mass. Although we can account for how the ejecta velocities scale with impact speed, the *survivability* of this material to actually return to the surface and ultimately migrate to the cold traps is nevertheless unknown. To understand how the amount of material delivered to the cold traps depends on impactor size, speed, and angle, such impacts may need to be individually simulated using DSMC, which is beyond the scope of this work. We therefore leave such studies to future projects.

## V. Summary and Conclusions

In this work, we use a technique, first used for studies of the Moon (*Stewart et al. 2011; Prem et al. 2015, 2018, 2019a, 2019b*), that combines hydrodynamic impact modeling (here the SOVA hydrocode) with the PLANET Direct Simulation Monte Carlo (DSMC) code to simulate the initial impact and aftermath of a 1 km radius comet striking Mercury's north pole at 30 km/s at a 60° angle. Although previous studies had made considerable progress modeling the migration of water across Mercury's surface, this effort is the first to model the full gas dynamic, photolytic and radiation transport evolution of an impact-generated vapor plume on Mercury from initial impact to ultimate fate, and is thus the first to account for the three dominant fates of these water molecules (ballistic escape, photodestruction, and migration to the cold traps).

To reduce computational expense, we scaled the state of an evolving impact plume from a previous study using the same impact scenario (albeit into the Moon), starting from 30 s post-impact. this lunar plume simulation was then placed at the surface of Mercury to continue the simulation, with negligible errors associated with this approach. PLANET DSMC then tracked the evolution of the plume as its individual molecules interacted with solar photons, Mercury, and other molecules. PLANET DSMC then recorded the fates of these molecules as they photodestructed, escaped the system, or landed in a cold trap; PLANET DSMC also recorded regular snapshots of the bulk (statistical) atmosphere/exosphere.

We find that the evolution of the resulting water vapor plume could be separated into four distinct phases: early plume, reentry, quasi-steady, and late phases. Gas dynamics in the earliest stages of plume evolution accelerate the vapor within, causing a greater fraction of molecules to escape Mercury's gravity than would be expected from the speeds of the gas molecules at the start of the simulation alone; this gas-driven acceleration increased the fraction of ballistically escaping molecules by nearly 20%. Later, as the plume expands toward its maximum extent before collapsing onto itself, we find that atmospheric self-shielding dramatically increases the survival probability of water vapor molecules against photodestruction, as only 0.18% of water molecules would be expected to survive 24 hours of exposure to sunlight at Mercury's distance from the Sun without it. By accounting for the absorption of EUV photons in the sunward side of the plume, we find that the fraction of molecules that survive through to later phases is dramatically increased.

As this plume collapses during the reentry phase, a reentry shock forms when water vapor molecules fall back to the surface of Mercury after the initial expansion of the plume. This shock layer is hot (~1000 K) and compresses the atmosphere below, creating a dense (~$10^{18}$ molecules/m$^3$) surface atmosphere that is largely shielded from photodestruction. When present, this surface atmosphere guides molecules to the terminator where they adsorb onto the surface, forming a relatively thick frost layer on the night side of Mercury. A more tenuous columnar shock structure formed at the antipode of the impact site.

After the shock atmosphere dissipates at the conclusion of the reentry phase, the exosphere is populated instead by surface frosts slowly rotating into daylight. This produces a narrow Dawn Atmospheric Enhancement (DAE) as adsorbed water on Mercury's surface rotates into daylight and desorb. This DAE is a key feature of the quasi-steady phase, and is responsible for much of the water that migrates to the cold traps. Ultimately, we find that 65% of the initial water molecules escaped the Mercury system to the Hill sphere, of which ~80% photodestruct prior to reaching the Hill sphere. A further 23 % of the initial molecules photodestruct while bound to Mercury. In the end, 14% of the initial water molecules from the impact migrated to the cold traps, forming deposits equivalent to ~1-10 mm of water ice. These relatively thick deposits are enabled by self-shielding against photodestruction.

This study provides an opportunity for comparing the effects of this impact into Mercury with identical simulations of an identical impact into the Moon (*Prem et al. 2015; 2019*), whose lower gravity and weaker solar radiation environment result in some important differences in the evolution of the vapor plume and the fate of the water molecules. At 30 s post-impact, only 28% of the water molecules were gravitationally bound to the Moon, compared to 45.5% for Mercury, due to their different escape speeds. Surprisingly, little of this additionally retained mass survives past the initial evolution of the plume due to the enhanced photodestruction environment at Mercury. While both Mercury and the Moon develop a reentry shock atmosphere with similar densities (~$10^{18}$ molecules/m$^3$ at the surface and ~$10^{17}$ molecules/m$^3$ at the shock layer) and structures, Mercury's shock atmosphere is more compressed due to its stronger gravity: Mercury's atmospheric scale height is ~40 km, about half that of the Moon's atmosphere (85 km). This difference in scale height results in a more compressed atmosphere on Mercury, with a height of ~86 km compared to ~200 km on the Moon. Similarly, the DAE on Mercury is narrower than on the Moon due to Mercury's higher gravity. In the end, 14% of the initial water molecules from the impact migrated to Mercury's cold traps, which is significantly higher than the estimated ~5% for the Moon. This difference is likely due to Mercury's larger cold trap area and higher gravity.

Unlike on the Moon and Mercury, negligible water vapor from a 30 km/s comet impact would be retained by Ceres' and Vesta's paltry gravity, amounting to, at most, a cold trapped deposit only a few molecules thick on Ceres. This highlights the importance of gravity in the delivery and retention of water from impact events.

## VI. Acknowledgments

We would like to thank Nancy Chabot and Elizabeth McFarland for identifying and providing the cold traps used in this work. We also would like to thank Jennifer Hanley, for useful discussions regarding rovibrational spectroscopy. We wish to acknowledge the Texas Advanced Computing Center (TACC), who hosts, maintains, and supports the Stampede2 and Lonestar6 supercomputing clusters, with which we ran the PLANET DSMC code. This work was supported by NASA award 80NSSC17K0725.

## VII. Open Research

*Data Availability Statement*

Our code, including all input files that we used to run our code and generate our results are available on the Texas Data Repository; this includes our branch of the PLANET DSMC code used to create this work (all C code files), input files, and input files from *Prem et al. (2015)* 30 s post-impact. These resources are available online for download, see *Steckloff et al. (2026)* or https://doi.org/10.18738/T8/LZ9KOU,

## VIII. Conflict of Interest Statement

The authors are unaware of any potential, or apparent, conflicts of interest with this research.

## IX. Appendix

The cold traps in our simulation assume a known location (latitude and longitude), circular shape with a known radius, and trapping efficiency. For discrete cold traps (i.e., known locations of permanent shadow), this trapping efficiency is 100%. To account for small and sub-resolution cold traps, our simulation allows for latitude-dependent bands with trapping efficiencies less than 100%; in essence, these trapping efficiencies denote the fraction of that band's area that comprise small cold traps (each with 100% trapping efficiency). To incorporate such bands into our simulation, we center the cold trap on either the North or South Pole and specify their equivalent radii along the surface (assuming spherical cap geometry) and trapping efficiency. We obtain these values (see Tables A1 and A2) from *Deutsch et al. 2016* and *Chabot et al. 2018*).

| Cold Trap ID (ref) | Latitude (deg) | Longitude (deg) | Radius (km) | Trapping Efficiency (%) |
|---|---|---|---|---|
| NP 3 (a) | 87.76 | 79.56 | 21.62 | 100 |
| NP 2 (a) | 88.8 | 149.27 | 21.07 | 100 |
| NP 1 (a) | 88.48 | 233.65 | 15.98 | 100 |
| NP 10 | 85.95 | 319.55 | 13.54 | 100 |
| NP 5 | 89.53 | 183.29 | 11.56 | 100 |
| NP 11 | 86.04 | 227.55 | 11.25 | 100 |
| NP 18 | 85.85 | 188.3 | 10.87 | 100 |
| NP 85 | 84.76 | 64.12 | 10.66 | 100 |
| NP 99 | 84.87 | 353.75 | 9.12 | 100 |
| NP 72 | 82.66 | 110.16 | 9.01 | 100 |
| NP 84 | 84.81 | 59.21 | 8.98 | 100 |
| NP 105 | 82.54 | 317.34 | 8.81 | 100 |
| NP 145 | 82.77 | 208.77 | 8.55 | 100 |
| NP 101 | 83.18 | 310.2 | 8.54 | 100 |
| NP 147 | 81.72 | 207.72 | 8.54 | 100 |
| NP 104 | 82.23 | 342.49 | 7.94 | 100 |
| NP 12 | 87.65 | 211.47 | 7.93 | 100 |
| NP 87 | 84.8 | 68.69 | 7.63 | 100 |
| NP 186 | 80.45 | 142.96 | 7.49 | 100 |
| NP 190 | 78.45 | 204.78 | 7.42 | 100 |
| NP 80-90 deg | 90 | 0 | 425.81 | 1.37 |
| NP 70-80 deg | 90 | 0 | 851.62 | 0.32 |
| NP 67-70 deg | 90 | 0 | 979.36 | 0.04 |

**Table A1:** *List of major north polar cold traps on Mercury, from (Deutsch et al. 2016; Chabot et a. 2018).* All cold traps are assumed to be circular except for the diffuse cold traps defined by latitude bands, which may result in a “donut” shape; trapping efficiency lists the probability that an incident molecule is cold trapped.

| Cold Trap ID (ref) | Latitude (deg) | Longitude (deg) | Radius (km) | Trapping Efficiency (%) |
|---|---|---|---|---|
| SP 271 | -86.45 | 69.33 | 23.03 | 100 |
| SP 184 | -87.5 | 202.48 | 20.11 | 100 |
| SP 270 | -87.12 | 352.97 | 19.79 | 100 |
| SP 152 | -86.2 | 286.1 | 19.49 | 100 |
| SP 183 | -87.5 | 222.44 | 15.99 | 100 |
| SP 388 | -81.92 | 81.39 | 14.16 | 100 |
| SP 81 | -87.6 | 296.51 | 13.60 | 100 |
| SP 185 | -87.84 | 184.84 | 12.61 | 100 |
| SP 186 | -87.46 | 173.88 | 12.32 | 100 |
| SP 275 | -84.8 | 69.04 | 12.26 | 100 |
| SP 282 | -84.2 | 81.58 | 12.02 | 100 |
| SP 246 | -87.2 | 277.73 | 11.25 | 100 |

| SP 284 | -83.97 | 75.33 | 11.06 | 100 |
|---|---|---|---|---|
| SP 310 | -84.86 | 345.84 | 10.75 | 100 |
| SP 318 | -85.44 | 17.01 | 10.70 | 100 |
| SP 375 | -81.16 | 152.17 | 10.56 | 100 |
| SP 343 | -83.68 | 292.8 | 10.18 | 100 |
| SP 198 | -83.75 | 160.08 | 9.65 | 100 |
| SP 392 | -80.19 | 77.32 | 9.62 | 100 |
| SP 354 | -82.63 | 290.8 | 9.54 | 100 |
| SP 80-90 deg | -90 | 0 | 425.81 | 3.17 |
| SP 73-80 deg | -90 | 0 | 723.87 | 0.37 |

**Table A2:** *List of major south polar cold traps on Mercury, from (Deutsch et al. 2016; Chabot et al. 2018).* All cold traps are assumed to be circular; trapping efficiency lists the probability that an incident molecule is cold trapped.